%% file: main.tex
\documentclass[fleqn,10pt]{olplainarticle}

\usepackage{xfrac}
\usepackage{graphicx}
\usepackage{xspace}
\usepackage{hyperref}
\usepackage{bm}
\usepackage{{booktabs}}

\newcommand{\nuc}[3]{\ensuremath{^{#2}_{#3}}\text{#1}\xspace}
\newcommand{\cs}{\nuc{Cs}{137}{}}
\newcommand{\cf}{\nuc{Cf}{252}{}}
\newcommand{\na}{\nuc{Na}{22}{}}
\newcommand{\co}{\nuc{Co}{60}{}}
\newcommand{\li}{\nuc{Li}{6}{}}
\newcommand{\ba}{\nuc{Ba}{133}{}}
\newcommand{\am}{\nuc{Am}{241}{}}

\newcommand{\uCi}{{\textmu}Ci\xspace}

\DeclareMathOperator*{\argmin}{\arg\!\min}

\title{Demonstration of a new CLLBC-based gamma- and neutron-sensitive free-moving omnidirectional imaging detector}

\author[1,*]{Jayson R.~Vavrek}
\author[1]{Ryan Pavlovsky}
\author[1]{Victor Negut}
\author[1]{Daniel Hellfeld}
\author[1]{Tenzing H.Y.~Joshi}
\author[1]{Brian J.~Quiter}
\author[1]{Joshua W.~Cates}
\affil[1]{Applied Nuclear Physics Program, Nuclear Science Division, Lawrence Berkeley National Laboratory, 1 Cyclotron Road, Berkeley, CA, USA, 94720}
\affil[*]{Corresponding author: \href{mailto:jvavrek@lbl.gov}{jvavrek@lbl.gov}}


\begin{abstract}
We have developed a CLLBC-based gamma- and neutron-sensitive multi-channel omnidirectional imaging detector, suitable for handheld or vehicle-borne operation and capable of quantitative radiation mapping in 3D.
The system comprises 62 CLLBC modules in an active-masked configuration, and is coupled to a Localization and Mapping Platform (LAMP) suite of contextual sensors that provides a 3D map of the environment.
The contextual and radiation data is combined using Scene Data Fusion (SDF) methods to better inform the reconstruction of the source radiation distribution from variations in the measured counts as the detector moves throughout the 3D environment.
Here, we first present benchtop-scale characterization studies for both the neutron and gamma ray channels.
In tandem, we present Geant4 simulations of both the single-crystal and full-system detection efficiencies over the omnidirectional field of view, and compare against validation measurements.
We then demonstrate the imager's capabilities in a variety of different scenarios, ranging from free-moving handheld simultaneous measurements of \cs and \cf to more challenging motion-constrained or static measurement scenarios.
In several of these scenarios we also demonstrate how the full omnidirectional multi-crystal responses markedly improve the reconstruction quality.
The imager is therefore a promising system for conducting simultaneous gamma and neutron radiation measurements in applications such as homeland security, contamination mapping, and nuclear decommissioning.
\end{abstract}

\begin{document}

\flushbottom
\maketitle
\thispagestyle{empty}

\section{Introduction}
Quantitative radiation imaging aims to reconstruct both the intensity and spatial location or distribution of ionizing radiation produced by radioactive material in an environment.
Over the past decade, the scene-data fusion (SDF) technology developed at Lawrence Berkeley National Laboratory (LBNL) and the University of California, Berkeley (UCB), has advanced quantitative radiation imaging by coupling radiation measurements with models of the measured environment, thereby enabling the attribution of radiation intensities to specific locations or features in the scene~\cite{vetter2019advances}.
These measurements are typically made using radiation detectors on free-moving platforms such as small unmanned aerial systems (UAS), ground vehicles, or human operators, in order to achieve better spatial coverage and thus confidence in both the scene model and the radiation image.
SDF has been demonstrated in a wide variety of applications, including contamination mapping in indoor laboratories~\cite{haefner2017handheld, hellfeld2021free}, controlled field tests~\cite{vavrek2024surrogateIII}, forested wetlands~\cite{vavrek2023free}, and accident sites such as Fukushima and Chornobyl~\cite{vetter2019advances}; urban radiological search~\cite{bandstra2016radmap, salathe2021determining}; and mapping naturally-occurring \nuc{K}{40}{}~\cite{joshi2017measurement}.
Recent work has extended the SDF concept to the mapping of minimum detectable activity~\cite{bandstra2022mapping} and to including attenuation from the scene model~\cite{bandstra2021improved}.
Outside of LBNL/UCB, a number of groups have recently developed similar gamma-ray imaging concepts/systems, typically using less scene information, including using deep neural networks~\cite{okabe2024tetris}, coded apertures~\cite{boardman2020single}, or active coded masks~\cite{ghelman2024wide} for source direction localization; particle filters for multiple point source reconstruction with attenuation in 3D~\cite{kemp2023real}; high-angular-contrast detector designs~\cite{kitayama2022feasibility, hu2023wide}; simultaneous detection of gammas and neutrons~\cite{rossi2023gamma} for 2D imaging~\cite{steinberger2020imaging, lopez2022neutron}; and more traditional Compton cameras~\cite{sinclair2014silicon, sinclair2020end, murtha2021tomographic}.
Finally, commercial imaging/mapping systems leveraging SDF or similar technology have recently become available from Gamma Reality Inc.~\cite{gri_lamp} and H3D Inc.~\cite{h3d_products}, and have been adopted for routine monitoring tasks in several nuclear power plants.

The current state-of-the-art of SDF-enabled detector systems use the Localization and Mapping Platform (LAMP)~\cite{Pavlovsky2018}, a suite of contextual sensors consisting of a light detection and ranging (LiDAR) unit, inertial measurement unit (IMU), on-board computer, and optionally a video camera.
Scene models and detector trajectories in the scene are found using LiDAR SLAM (simultaneous localization and mapping)~\cite{Durrant-Whyte2006_1, Durrant-Whyte2006_2, Hess2016}, which couples LiDAR scans of the environment with the IMU estimates of the detector's position and orientation (together ``pose'') in 3D space.
The 3D model enables fully 3D reconstructions, 2D reconstructions confined to (for instance) a ground plane, or ``2.5D'' reconstructions confined to the collection of visible surfaces in the LiDAR model of the environment.
Because in most imaging applications radionuclides are expected to be present on surfaces (ground, walls, tables, etc.) rather than in mid-air, the 2.5D reconstruction modality both greatly speeds up reconstruction performance and effectively fuses scene and radiation information to provide a scene-aware reconstruction.

Given a quantitative detector response function (i.e., detection efficiency vs incident direction), SDF employs reconstruction algorithms such as maximum-likelihood expectation-maximization (ML-EM)~\cite{shepp1982maximum}, regularized ML-EM or maximum \textit{a posteriori} expectation-maximization (MAP-EM), gridded point source likelihood (GPSL), additive point source likelihood (APSL)~\cite{hellfeld2019gamma, vavrek2020reconstructing}, or Compton-imaging versions thereof~\cite{hellfeld2021free}, in order to find the most likely radiation distribution giving rise to the measured counts.
The response function is typically computed through Monte Carlo simulations using frameworks such as Geant4~\cite{agostinelli2003geant4, allison2006geant4, allison2016recent} and is often expressed in terms of the ``effective area'' $\eta$, which comprises both geometric and intrinsic detection efficiency (see Section~\ref{sec:response_benchmarking}).
Several features in a given detector response will influence its imaging performance.
For gamma ray imaging, the detector material should have a high density $\rho$ and effective atomic number $Z_\text{eff}$ in order to provide a high stopping power and thus intrinsic detection efficiency.
The ability to modulate an anisotropic detection efficiency is also useful in breaking degeneracies inherent to the source reconstruction problem (see Section~\ref{sec:math}), especially if the detector is position-sensitive, i.e., can determine where in its active volume a given photon interaction occurred.
The detector energy resolution should be sufficiently narrow so that it can isolate photopeaks from various radionuclides.
Finally, to enable neutron imaging, the detector material must also be neutron-sensitive, and the neutron signature must be distinguishable from photon interactions.

In this work, we present a new SDF-capable imaging detector designed to take advantage of the above features.
The detector is dual-modality---sensitive to both gammas and neutrons---through its use of Cs$_2$LiLaBr$_{6-x}$Cl$_x$:Ce$^{3+}$ (CLLBC, RMD Inc.~\cite{rmd_cllbc_spec, guss2014scintillation}), which is an inorganic scintillator enriched in the neutron-sensitive nuclide \li.
Thermal neutron interactions with \li manifest as a peak at ${\sim}3$~MeVee (MeV electron equivalent), meaning they can often be reliably discriminated from photons based on pulse height alone.
CLLBC has a density of $\rho \simeq 4$~g/cm$^3$, between that of sodium iodide (NaI(Tl), $3.67$~g/cm$^3$~\cite[p.238]{knoll2010radiation}) and CdZnTe ($5.8$~g/cm$^3$), providing strong intrinsic detection efficiency.
The energy resolution of CLLBC is $3$--$4\%$ FWHM at $662$~keV~\cite{rmd_cllbc_spec}, among the best available of commercial inorganic scintillators, and typically sufficient for isolating a given photopeak.
The system uses $62$ separate $\sfrac{1}{2}"$ ($1.27$~cm) CLLBC cubes, each of which is individually read out, thereby providing position sensitivity with the resolution of the crystal size.
Importantly, the CLLBC cubes are arranged in a quasi-random, partially-populated stack of four $6 \times 6$~grids---this active-masked pattern, along with the LAMP contextual sensors and battery nearby the detectors, provides a strongly-anisotropic pattern in the per-detector-module $4\pi$ angular detector responses.
While this quasi-random detector arrangement is also highly suitable for Compton imaging, in this work we focus on imaging based on singles events.
Finally, since the detector is coupled to a LiDAR SLAM system, it can localize and quantify both gamma and neutron sources in 3D.

In Section~\ref{sec:methods}, we give a more in-depth system description, show the benchtop-level performance of the system in terms of energy resolution and time-to-next-event (TTNE) histograms, and discuss simulations of the detector response.
We also provide an overview of the mathematical framework and reconstruction methods of quantitative imaging that will be used to characterize the imager's performance.
In Section~\ref{sec:results} we explore the results of the detector response simulations before moving on to the results of several laboratory and field demonstrations of the detector.
These demonstrations range from handheld surveys of indoor laboratories to outdoor vehicle-mounted surveys to various constrained measurement scenarios designed to showcase the efficacy of the omnidirectional anisotropic multi-crystal response.
Section~\ref{sec:discussion} then provides an additional discussion of the results and recommendations for future work.

\section{Methods and Materials}\label{sec:methods}

\subsection{System description}

\begin{figure}[!htbp]
    \centering
    \includegraphics[width=1.0\columnwidth]{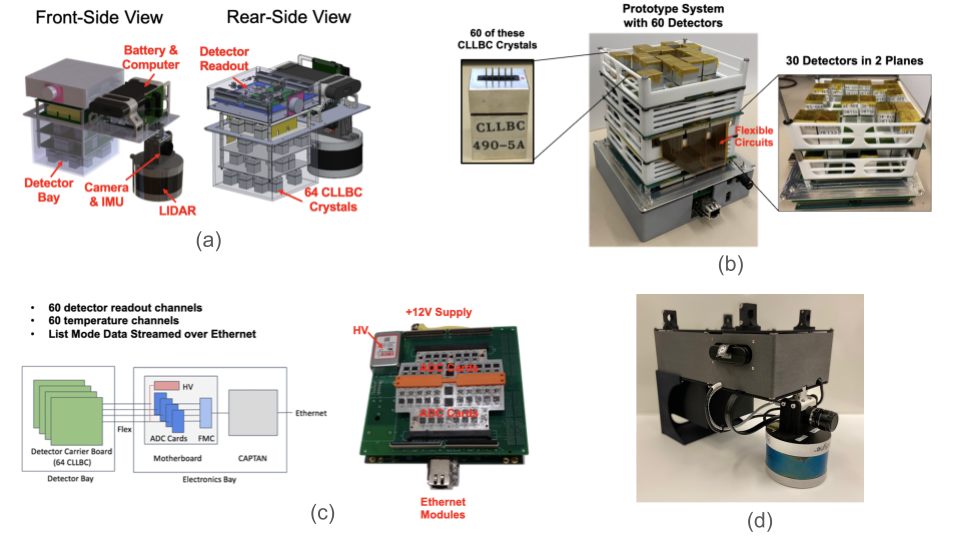}
    \caption{
        Overview of the multi-channel imager.
        (a) 3D design models of the system from the front and rear, showing the major components.
        (b) CLLBC detector modules arranged on circuit boards into a 3D coded or active masked array, upside-down compared to their orientation in the final system.
        (c) Signal processing overview from detector carrier boards to data out via Ethernet.
        (d) Photo of the assembled system with mounts for a UAS.
    }
    \label{fig:ngv2_overview}
\end{figure}

The imaging detector---see the overview in Fig.~\ref{fig:ngv2_overview}(a)---uses an array of Radiation Monitoring Devices Inc.\ (RMD) Cs$_2$LiLaBr$_{6-x}$Cl$_x$:Ce$^{3+}$ (CLLBC) crystals~\cite{rmd_cllbc_spec, guss2014scintillation} to enable both gamma and neutron detection.
The Li component is enriched from a natural \li abundance of ${\sim}7.5\%$ to $95\%$, since the thermal neutron detection channel relies on the inelastic reaction
\begin{align}
    \nuc{Li}{6}{3} \,+\, \nuc{n}{1}{0} \,\to\, \nuc{H}{3}{1} \,+\, \nuc{$\alpha$}{4}{2} \,+\, 4.78\,\text{MeV}.
\end{align}
The $4.78$~MeV of energy liberated in the reaction is split between the kinetic energies of the resulting triton and alpha, which ionize the medium and create a detectable scintillation pulse with an electron-equivalent energy of ${\sim}3.1$~MeV.
Photons are also detected through scintillation, with CLLBC achieving energy resolutions of $3$--$4\%$ full width half maximum (FWHM) at $662$~keV.

Each $\sfrac{1}{2}'' \times \sfrac{1}{2}'' \times \sfrac{1}{2}''$ CLLBC crystal (Fig.~\ref{fig:ngv2_overview}(b)) is coupled to an OnSemi ARRAYJ-60035-4P silicon photomultiplier (SiPM) array.
Anode signals from each detector are routed over flexible printed circuit boards (Fig.~\ref{fig:ngv2_overview}(c)) to data acquisition external to the detector bay (Fig.~\ref{fig:ngv2_overview}(d)).
Detector signals are concentrated onto four $16$-channel analog-to-digital conversion cards (Fig.~\ref{fig:ngv2_overview}(e)), comprising peak detect circuits with digitally controlled reset.
Digitized events are then packaged into individually timestamped, list-mode format by a shared field-programmable gate array (FPGA). 
The $J=62$ detector modules are arranged quasi-randomly in four horizontal planes of $14$--$16$ modules each; this quasi-random pattern was first implemented on the MiniPRISM system~\cite{pavlovsky2019miniprism} to provide an active-masked response, high detection efficiency in all directions, and large inter-crystal distances for Compton imaging---see also the later Fig.~\ref{fig:geom_geant4}.
We note that while the system was originally designed with $J=64$ modules, $J=62$ modules were installed and included in the simulations of Section~\ref{sec:response_benchmarking}, and often only $J=56$--$58$ read out data in this work's later measurements.
This module arrangement leads to a per-crystal anisotropic detector response $\eta(\theta, \phi)$ in two ways.
First, the detectors are partially occluded by the LAMP sensor suite (LiDAR, battery, etc.) in the forward and upward directions.
More importantly, however, the individual detector modules occlude each other, creating an active-masked detector.
The active-masked design obviates the need for heavy lead or tungsten passive coded apertures, which could achieve similar directionality to active-masked arrangements at the expense of reducing detection efficiency.
The active mask can therefore be thought of as a passive mask that is made from other detectors, in contrast to designs using passive volumes of W, Pb~\cite{okabe2024tetris, kitayama2022feasibility} or NaCl~\cite{farzanehpoor2021feasibility}.

The LAMP sensor suite itself~\cite{Pavlovsky2018, pavlovsky20193d} consists of an Intel NUC computer, a FLIR video camera, a Velodyne Puck LITE LiDAR unit, and a VectorNav IMU.
The LiDAR and IMU are used to perform LiDAR SLAM via Cartographer~\cite{Hess2016}, which can be run on the NUC in near-real-time at lower spatial fidelity or offline at higher fidelity, typically giving the detector position and orientation within the 3D LiDAR map with a time binning of $0.1$~s.
The radiation data, conversely, is stored as timestamped listmode data.
The system is powered by an Inspired Energy lithium ion battery pack, and is controlled and read out using the Robot Operating System (ROS) toolkit.
The LAMP body is a carbon fiber / nylon blend and supports mount points for a UAS or a handle for hand-carrying.
The completed system mass is $5.1$~kg, about $0.5$~kg of which is CLLBC.

\subsection{Mathematical framework}\label{sec:math}
We begin by considering an isotropic radiation detector with effective area $\eta$ making a single measurement with dwell time $t$ of a point source of radiation with intensity $w$ separated from the detector by a vector $\vec{r}$.
The mean number of counts expected during the measurement is, ignoring background and air attenuation,
\begin{align}\label{eq:lambda}
    \lambda = \frac{\eta w t}{4\pi |\vec{r}|^2}
\end{align}
and the measured counts are distributed according to Poisson statistics:
\begin{align}\label{eq:n}
    n \sim \text{Poisson}(\lambda).
\end{align}
Often we will be interested in the source reconstruction problem, inverting Eq.~\ref{eq:lambda} to determine the 3D position $\vec{r}$ and/or intensity $w$ of an unknown point source.
It is important to note that in this case, the isotropic Eq.~\ref{eq:lambda} is degenerate in two ways.
First, the intensity $w$ and squared distance $|\vec{r}|^2$ are collinear---for a single measurement, Eq.~\ref{eq:lambda} is sensitive only to their ratio.
Second, the distance vector $\vec{r}$ itself is degenerate---for a given $\lambda$ and $w$ there is a sphere of points at radius $|\vec{r}|$ that will satisfy Eq.~\ref{eq:lambda}.
Fortunately, the degeneracies of Eq.~\ref{eq:lambda} can be broken by making multiple independent measurements $i=1, \ldots, I$ (where $\lambda \to \lambda_i$, $t \to t_i$, etc.) with different distance vectors $\vec{r}_i$.
In the fully general case with multiple measurements and multiple source points $k=1, \ldots, K$, Eq.~\ref{eq:lambda} generalizes to
\begin{align}
    \label{eq:lambda_arr}
    \boldsymbol{\lambda} &= \boldsymbol{V} \boldsymbol{w}\\
    \label{eq:n_arr}
    \boldsymbol{n} &\sim \text{Poisson}(\boldsymbol{\lambda})
\end{align}
where $\boldsymbol{\lambda}$ is the vector of all $\lambda_i$, $\boldsymbol{n}$ is the vector of all $n_i$, $\boldsymbol{w}$ is the vector of all $w_k$, and $\boldsymbol{V}$ is the $I \times K$ ``system matrix'' with elements
\begin{align}\label{eq:sys_mat}
    V_{ik} = \frac{\eta_{ik} t_i}{4\pi |\vec{r}_{ik}|^2}.
\end{align}
The ``sensitivity'' to point~$k$ is then defined as
\begin{align}
    \varsigma_k \equiv \sum_i V_{ik},
\end{align}
which has dimensions of time and gives the expected number of counts per unit emission rate from point $k$ over the entire measurement.
The model is often further generalized to $j=1, \ldots, J$ independently-read-out detector modules.
Rather than form a 3D system matrix $V_{ijk}$, we can flatten indices via lexicographical ordering $(i, j) \to i + I(j-1)$ and therefore $I \to I J$.
For simplicity we work with $J=1$ in this section, but later we will indicate $J>1$ when necessary.

Reconstruction degeneracies can be even more strongly broken by additionally using a highly-anisotropic detector response $\eta(\theta, \phi)$, where $\theta$ and $\phi$ are the polar and azimuthal angles, respectively, of the source point in the detector's coordinate frame.
In practice, and as we will demonstrate in Section~\ref{sec:results}, this means free-moving detectors with highly-anisotropic response functions should offer improved imaging performance compared to static and/or isotropic detectors.

\subsection{Quantitative reconstruction methods}

Here we provide an overview of quantitative radiological image reconstruction methods.
For a more comprehensive treatment of these algorithms, we refer the readers to the Methods section of Ref.~\cite{hellfeld2021free} and references therein.

In general, SDF seeks the most likely source image $(\hat{\boldsymbol{w}}, \hat{\vec{\boldsymbol{r}}})$ from Eqs.~\ref{eq:lambda_arr}--\ref{eq:n_arr} by formulating the minimization problem
\begin{align}\label{eq:argmin}
    \hat{\boldsymbol{w}}, \hat{\vec{\boldsymbol{r}}} = \argmin_{\boldsymbol{w},\, \vec{\boldsymbol{r}}}\, \left[ \ell(\boldsymbol{n} | \boldsymbol{\lambda}(\boldsymbol{w}, \vec{\boldsymbol{r}})) + \beta R(\boldsymbol{w}) \right]
\end{align}
where the (negative) Poisson log loss is
\begin{align}
    \ell(\boldsymbol{n} | \boldsymbol{\lambda}) = \sum_{i=1}^I \left[ \lambda_i - n_i \log \lambda_i + \log \Gamma(n_i + 1) \right]
\end{align}
and $R(\boldsymbol{w})$ is an optional image regularization term with strength $\beta$.
Common regularizers include the sparsity-promoting $\ell_0$, $\text{log}$, and $\Gamma$ priors as discussed in Refs.~\cite[Sec.~II.C]{hellfeld2019gamma} and~\cite{lingenfelter2009sparsity}, and more recently, the $\ell_{\sfrac{1}{2}}$ regularizer~\cite{xu2010lhalf}.

For one or a small number of unknown discrete point sources, Eq.~\ref{eq:argmin} is written with separate $\boldsymbol{w}$ and~$\vec{\boldsymbol{r}}$ terms, typically uses no regularizer, and can be solved with methods such as Point Source Likelihood (PSL)~\cite{hellfeld2019gamma} or its multi-source extension Additive Point Source Localization (APSL)~\cite{hellfeld2019gamma, vavrek2020reconstructing}.
Similarly, the recently-developed generalized APSL (gAPSL)~\cite{greiff2021gamma}, which permits non-point-source kernels such as multivariate Gaussian distributions, uses a similar minimization that also includes source shape parameters.
In this work, most reconstructions use a discrete version of PSL, Gridded Point Source Localization (GPSL).

By contrast, in distributed source reconstructions, it is more common to define a voxellized grid in space and allow all (or a fixed subset of) voxels to contribute to the image; in this case the source positions $\boldsymbol{\vec{r}}$ are fixed, and maximum likelihood expectation maximization (ML-EM) or its regularized ($\beta > 0$) analog maximum \textit{a posteriori} expectation maximization (MAP-EM) are used to solve for the most likely source activities:
\begin{align}\label{eq:argmin_w}
    \hat{\boldsymbol{w}}(\boldsymbol{\vec{r}}) = \argmin_{\boldsymbol{w}} \; \left[ \ell(\boldsymbol{n} | \boldsymbol{\lambda}(\boldsymbol{w})) + \beta R(\boldsymbol{w}) \right].
\end{align}
The ML-EM ($\beta = 0$) solution to Eq.~\ref{eq:argmin_w} is the iterative update rule~\cite{shepp1982maximum}
\begin{align}\label{eq:mlem}
    \hat{\boldsymbol{w}}^{(m+1)} = \frac{\hat{\boldsymbol{w}}^{(m)}}{\boldsymbol{\varsigma}} \odot \boldsymbol{V}^\top \frac{\boldsymbol{n}}{\boldsymbol{V} \hat{\boldsymbol{w}}^{(m)}},
\end{align}
where $\odot$ denotes element-wise multiplication and the initial $\hat{\boldsymbol{w}}^{(0)}$ is often set to the flat image that produces the same total number of modeled counts as observed in the data, $\boldsymbol{n} \cdot \boldsymbol{1}$.
For the MAP-EM ($\beta > 0$) solution, we again refer the reader to Ref.~\cite{hellfeld2021free}.
There are a number of possible stopping criteria to avoid overfitting to noise as $m \to \infty$ (e.g., \cite{bissantz2008statistical, montgomery2020novel, ben2013heuristic}), but we adopt the simplest criterion of a fixed number of iterations (usually $10$--$100$).
These reconstruction algorithms can be applied to either `singles' data, where the incident photon deposits all of its energy in a single detector crystal, `doubles' data, where the incident photon first Compton scatters in one crystal before depositing the rest of its energy in another, or a combination thereof.
In this work, we focus on singles analyses, though Compton reconstructions could be explored in the future.

Our reconstruction algorithms are implemented in the Multi-modal Free-moving Data Fusion ({\tt mfdf}) package~\cite{joshi2020multi}, and make use of the LBNL-developed {\tt radkit} libraries~\cite{joshi2021radkit, hellfeld2021radkit} for handling radiological and scene data.
The entire system matrix (Eq.~\ref{eq:sys_mat}) is often too large to fit in computer memory, so we calculate each element $V_{ik}$ on the fly at each reconstruction iteration.
Both the system matrix calculation (Eq.~\ref{eq:sys_mat}) and the reconstruction (Eq.~\ref{eq:mlem}) are highly parallelizable, and thus are run on a graphics processing unit (GPU) via a PyOpenCL interface to the main Python3 analysis code.

\subsection{Detector response and benchmarking}\label{sec:response_benchmarking}

\begin{figure}[!htbp]
    \centering
    \includegraphics[width=0.79\columnwidth]{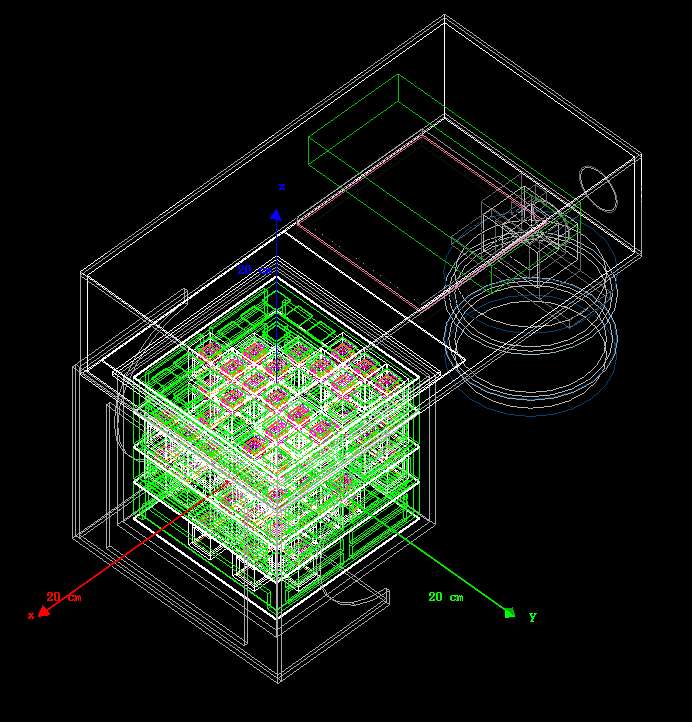}
    \caption{
        Geant4 model of the imaging detector, showing a rotated orthographic view of the entire system.
        The CLLBC crystals are white and are attached to red/orange SiPM components.
        The battery is the large dark green box (contained within the larger white LAMP box), while the lidar is the set of gray cylinders.
        The $20$~cm $xyz$ axes show the orientation and scale of the detector coordinate system.
        Despite the detector $x$-axis pointing away from the LiDAR, the detector-to-LiDAR direction is typically referred to as ``forward'' or the ``front'' for the overall system.
    }
    \label{fig:geom_geant4}
\end{figure}

Detector responses are measured with various sources and compared against Geant4~\cite{agostinelli2003geant4, allison2006geant4, allison2016recent} simulations.
Geometries are modelled to relatively high detail as shown in Fig.~\ref{fig:geom_geant4}---each simulated detector module includes the $\sfrac{1}{2}'' \times \sfrac{1}{2}'' \times \sfrac{1}{2}''$ CLLBC crystal (with $95\%$ \li abundance and $\rho = 4.0$~g/cm$^3$ overall density), aluminum module casing, silicon photomultiplier (SiPM), and various internal teflon wrap, printed circuit board (PCB), air gap, and other structural layers.
The modules are placed flush with $12 \times 12 \times 0.146$~cm PCBs and housed in a carbon fiber / nylon blend detector bay of thickness $2.5$~mm, which is attached to a simplified model of the LAMP box and sensor suite.

We conducted two studies benchmarking the imager's response in terms of its effective area~$\eta$.
In the first set of simulations, we use idealized source and world geometries to map out the effective area vs angle in $4\pi$ for both thermal neutrons and photons of various energies.
In the second, we developed a more realistic simulation of a \cf source measurement in order to better capture the complex environmental scatter and moderation.
A Python2.7-wrapped Geant4 v10.5 is used for the $4\pi$ photon and neutron simulations, while the standard C++ v10.7.p01 is used for the later \cf neutron validation simulations.

\subsubsection{Angular response simulations}\label{sec:4pi_sims}

The omnidirectional response simulations are used to generate the angular detector response maps $\eta(\theta, \phi)$ used in the reconstructions of Section~\ref{sec:results}, but are also used to compare against benchmarking measurements.

The primary particles simulated are either monoenergetic neutrons of $25$~meV kinetic energy to represent a thermal flux or photon beams at $59$~keV (\am), $186$~keV (\nuc{U}{235}{} series), $356$~keV (\ba), $662$~keV (\cs), $1001$~keV (\nuc{U}{238}{} series), $1332$~keV (\co), and additionally $100$~keV and $125$~keV to better capture the low-energy $\eta$ vs $E$ behavior.
The primaries are generated in a ``far-field'' configuration, i.e., a parallel circular beam of radius $R=15$~cm directed towards the origin of the CLLBC detector array, which itself lies in the center of the $6\times 6$ possible $xy$ module positions and the $z$ center of the middle PCB.
The origin of the beam is placed $1$~m away from the detector origin in order to include a small amount of attenuation and downscatter by air.
No other volumes are present in the simulation.
At each beam position, $N_p = 10^7$ primaries are generated in order to reduce statistical uncertainties.

To generate $4\pi$ angular responses, these gammas (neutrons) are generated at $3072$ ($768$) points on the $4\pi$ sky according to the HEALPix~\cite{gorski2005healpix, Zonca2019} {\tt nside=16} {\tt (8)} spherical pixelization schemes, which have average pixel separations of $3.7^\circ$ ($7.3^\circ$).
Given the discrete set of $3072$ ($768$) directions and (for gammas only) six energies, responses at intermediate angles and energies can be estimated through interpolation.
In the fixed-orientation case, effective area vs energy can be interpolated through standard linear-linear or log-log techniques.
In the fixed-energy case, effective area vs orientation is computed using HEALPix's spherical bilinear interpolation using the four nearest neighbor pixels.
We note that the angular interpolation can perform poorly in directions corresponding to the six faces of the detector array (which are often most convenient for validation measurements) for two compounding reasons.
First, aside from the lowest-resolution {\tt nside=1} pixel scheme, HEALPix maps do not place pixel centers in the six canonical up/down/left/right/fore/aft directions, so either interpolation or dedicated on-axis simulations are required.
Second, the angular response changes rapidly near the on-axis directions as the beam direction aligns with the crystal array.
Exactly on-axis, there are streaming paths (small air gaps and lower-$Z$ aluminum casing) accessible between the CLLBC modules that are obscured when the incident beam direction is rotated by even a few degrees.
To avoid this problem, additional simulations can be run with exactly on-axis beams.

The {\tt FTFP\_BERT\_HP\_LIV} physics list is used for improved low-energy physics accuracy for both neutron ({\tt HP}) and photon ({\tt LIV}) simulations.
Secondaries are tracked down to a production threshold of $10$~{\textmu}m, and information from any particle step within the CLLBC volumes is saved to disk for offline processing.
In the photon response analysis, energy deposition by photons and electron secondaries is summed for each event and each detector.
The summed energies are then blurred according to a Gaussian energy resolution model, and the number of events with energy within a $\pm 4\, \sigma$ window is used as the number of detected photons $N$.
In the neutron response analysis, the number of detected thermal neutrons~$N$ is determined by selecting on events with the {\tt neutronInelastic} process that produce both a triton and alpha particle.
We note that while much of the literature on CLLBC (e.g., Ref.~\cite{guss2014scintillation}) discusses ``neutron capture'' on \li, the {\tt ncapture} process in Geant4 refers specifically to neutron capture \textit{without} the inelastic production of heavy secondaries.
In both the gamma and neutron analyses, an $8.25$~mm radial acceptance is applied within each CLLBC crystal in order to model incomplete light collection from the corners, thereby reducing the active volume by $10.5\%$.
This heuristic is based on past modelling of CLLBC efficiencies in Geant4, and is discussed further in Section~\ref{sec:discussion}.

Responses are then computed in terms of the detector's effective area, which depends on the source-to-detector geometry and is found via the ratio of detected counts $N$ to the total particle fluence $\Phi$:
\begin{align}\label{eq:eff_area}
    \eta(\theta, \phi) = \frac{N}{\Phi} = \frac{N}{N_p} \cdot
    \begin{cases}
        \pi R^2, &\text{ far-field parallel circular beam source of radius $R$}\\
        4 \pi |\vec{r}|^2, &\text{ isotropic point source}.
    \end{cases}
\end{align}
Because we opted for far-field beams in the gamma response simulations, the modelled responses are independent of distance and thus are functions of angle alone, $\eta \equiv \eta(\theta, \phi)$, where $\theta$ is the polar angle from the detector $z$-axis and $\phi$ is the azimuthal angle.

\subsubsection{Angular response measurements}\label{sec:gamma_validation_measurements}
We conducted an initial set of response validation measurements via static dwells of ${\gtrsim}300$~s with calibration sources placed on-axis $2$~m away from the faces of the imager's detector bay.
\cs measurements were made for the left, right, aft, and fore (LiDAR-wards) faces.
\am and \co measurements were also taken at the imager's left face to measure the low- and high-energy response, respectively.
As described below, benchmarking of the thermal neutron response required further dedicated simulations.

Effective areas $\eta$ were then computed for each measurement using Eq.~\ref{eq:eff_area} under the isotropic point source case.
The number of primaries~$N_p$ during the irradiation was computed using the {\tt becquerel} library to perform decay corrections.
The number of counts~$N$ was determined by fitting the photopeak in the region of interest (ROI) for each nuclide with a Gaussian and a linear background term.
In the \cs and \co measurements, data from all detector crystals was used, while in the lower-energy \am measurements, only data from the first layer of crystals directly exposed to the source was used.
A fixed $|\vec{r}| = 2.07$~m to every crystal was assumed, except for the \am measurements, in which case $|\vec{r}| = 2.02$~m.
Air attenuation over these distances was also included in the model.
Corresponding simulated effective areas were computed using the simulation framework described in Section~\ref{sec:4pi_sims}, with dedicated on-axis simulations rather than interpolating between the nearest HEALPix coordinates.

As will be shown later, these initial validation measurements suffer from quite large systematic errors (especially in alignment) compared to the validation that can be achieved through other measurement modalities.
However, we include them here to help demonstrate the difficulty in quantitative validation methods for these types of imagers.

\subsubsection{\texorpdfstring{\cf}{Cf-252} validation simulations}\label{sec:cf252_validation_simulations}

As discussed below, we designed dedicated outdoor \cf measurement scenarios with a goal of easy-to-model environmental scattering.
In these outdoor scenarios, we model the earth, road surface, and top of the aluminum ladders below the imager and the source.
As opposed to the idealized far-field pure-thermal flux used in Section~\ref{sec:4pi_sims}, we simulate isotropic \cf decay directly, then allow the fast neutrons from fission to partially thermalize in a known moderator volume, escape from the moderator in all directions, travel through air, and then reach the imager.
The direct decay simulation requires Geant4 version 10.7.p01 for the G4RadioactiveDecay module, so this is developed as a standalone C++ simulation as opposed to the $4\pi$ simulation tool from above that uses Geant4 v10.5 and Python2.7.
We do however read in the imaging detector geometry from the $4\pi$~simulations using Geant4's GDML~\cite{chytracek2006geometry} modules to ensure consistency in geometry.

Accurate thermal neutron simulation in Geant4 requires the use of additional, non-default thermal elastic neutron scattering data libraries in addition to the use of the neutron high-precision ({\tt \_HP}) physics lists.
We activate these libraries in our \cf simulation according to~\cite[Section~5.2.3]{geant4_application_book}, and then replace all hydrogen in the detector materials with the specialized thermal scattering version, assuming polyethylene ({\tt TS\_H\_of\_Polyethylene}) is a representative chemical matrix for all hydrogenous detector materials. 
In addition, we replaced the polyethylene moderator with a (weakly-moderating) graphite cylinder with $2.75$~cm-thick walls (and used its corresponding thermal-library-enabled element, {\tt TS\_C\_of\_Graphite}), since graphite has generally shown better agreement in benchmarking studies than polyethylene (e.g., \cite{shin2014geant4}).
Finally, for computational performance, we discard the $97\%$ of \cf decays that result in $\alpha$ decay instead of spontaneous fission.

\subsubsection{\texorpdfstring{\cf}{Cf-252} validation measurements}

We initially conducted measurements similar to the gamma measurements of Section~\ref{sec:gamma_validation_measurements} with a polyethylene-moderated \cf source, but found that indoor neutron measurements suffered from substantial room return thermal scattering.
The indoor \cf measurements did not follow the inverse square law, for instance, due to thermal scattering off of the concrete walls and floor.
To avoid modeling complicated indoor structures in the validation simulations, we instead conducted measurements outdoors with both the imager and a $155$~\uCi \cf source on metal ladders about $2.5$~m off the ground.
The source and imager were placed ${\sim}1.5$~m apart and measurements were taken from the front, back, left, and right sides of the imager.
An additional measurement was made from below with the source on the ground, but since the top of the graphite cylinder was open, the imager was exposed to a larger fast flux as well.
Due to the influence of minor non-linearity in the response of the front-end signal processing chain, the global thermal neutron capture signature was broadened (see for example the later Fig.~\ref{fig:ngv2_cs137_walkaround}), so instead of performing a peak fit to extract the number of neutron counts, a simple integral of the $(2800, 3800)$~keV region was used instead.
To account for the non-negligible background observed during these measurements, we also fit an exponential background based on the $(2500, 2800)$~keV and $(3800, 4000)$~keV windows.
Subtracting this contribution resulted in a ${\sim}40\%$ decrease in the front, back, left, and right directions, and ${\sim}15\%$ in the bottom direction.

\section{Results}\label{sec:results}

Here we present demonstration results on the imager, starting with a benchtop characterization of its spectral and timing performance.
We then show its simulated responses, followed by a series of roughly increasingly-complex measurement scenarios.

\subsection{Benchtop characterization}

\begin{figure}[!htbp]
    \centering
    \includegraphics[width=1.0\columnwidth]{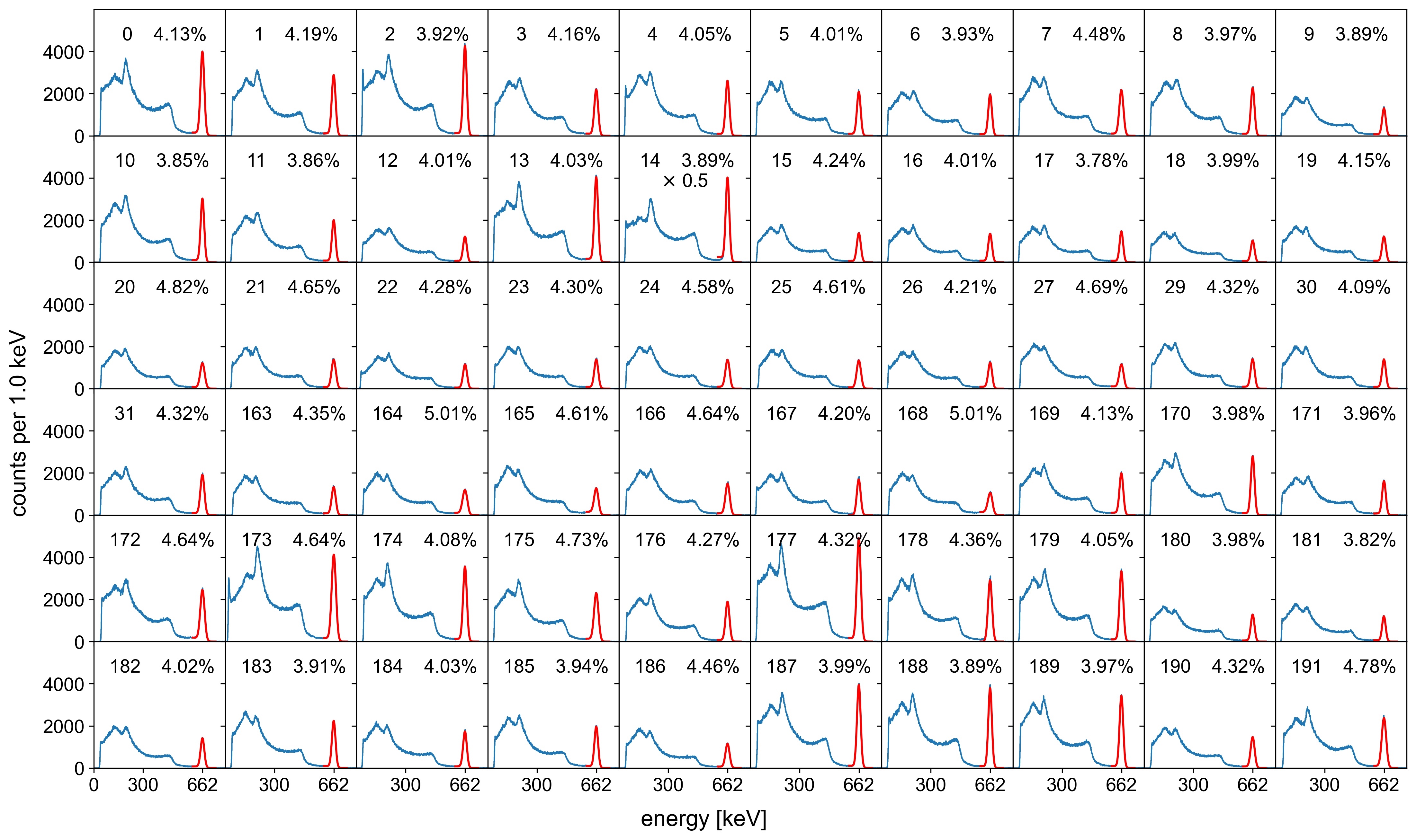}
    \caption{
        Individual \cs spectra (blue) and photopeak fits (red).
        The labels show the readout channel ID and the $661.7$~keV peak relative resolution (FWHM) for each detector module.
        The spectrum marked with ``$\times 0.5$'' was downsampled by a factor of $0.5$ in order to fit all spectra on the same $y$ scale, but the fit was computed with the full spectrum.
        Unlike the later Fig.~\ref{fig:resp_59keV}, the detectors are not grouped by horizontal plane.
    }
    \label{fig:individual_fits_Cs137}
\end{figure}

\begin{figure}[!htbp]
    \centering
    \includegraphics[width=1.0\columnwidth]{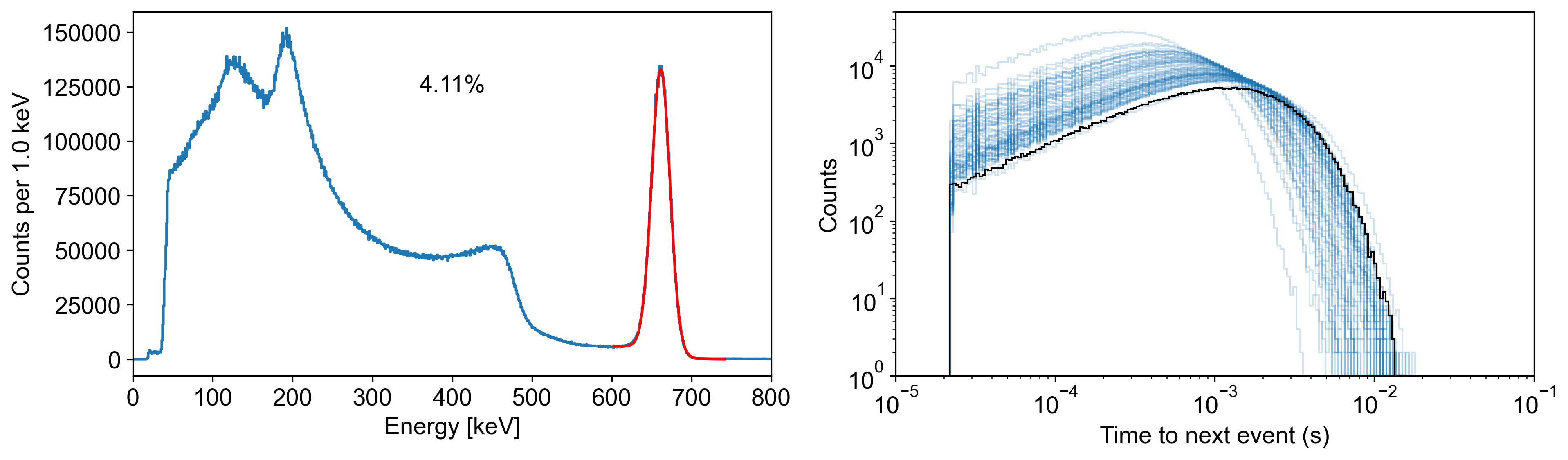}
    \caption{
        Left: global \cs spectrum and photopeak fit.
        Right: time-to-next-event (TTNE) histogram for each of the $60$ detector channels (blue), using $200$ log-spaced bins from $10^{-5}$ to $10^{-1}$~s.
        The black curve shows $3.3 \times 10^{5}$ synthetic samples of an exponential distribution with scale parameter $1.2 \times 10^{-3}$ for reference.
    }
    \label{fig:global_spec_indiv_ttne}
\end{figure}

The initial spectral performance characterization of the imager consisted of determining the resolution at the $661.7$~keV \cs photopeak for both the $60$ individual detector modules as well as the global spectrum.
Each module was first individually energy-calibrated using a fourth-order polynomial fit to the photopeak positions of a \nuc{Th}{228}{} spectrum, and the resulting calibrated spectrum was then blurred with a Gaussian kernel of $\sigma = 2$~keV to eliminate the resulting aliasing.
The imager was then irradiated with a \cs source, which was moved to irradiate multiple imager faces before being placed in a fixed location for a dwell time of ${\sim}400$~s.
The \cs photopeak in each module was then fit with a Gaussian peak shape plus error function background model, and resolutions were computed in terms of the full width half max (FWHM) relative to the centroid energy.
As shown in Fig.~\ref{fig:individual_fits_Cs137}, these FWHM values range from $3.78\%$ to $5.02\%$, and have a mean of $4.23\%$.
Fig.~\ref{fig:global_spec_indiv_ttne} then shows the same procedure applied to the global spectrum, where the global resolution is found to be $4.11\%$.
Fig.~\ref{fig:global_spec_indiv_ttne} also shows the time-to-next-event (TTNE) histogram for each module during the static dwell portion of the measurement, demonstrating consistency with the expected exponential wait time distribution.
Related, the radiation timestamps were found to exhibit a ${\sim}0.55$~s delay with respect to the trajectory timestamps in the MiniPRISM DAQ, which is identical to this imager's DAQ---see Ref.~\cite{vavrek2024surrogateII}.
As a result, we apply this small delay as part of the overall system calibration in the reconstructions of Section~\ref{sec:results}.

\subsection{Response simulation results}
\begin{figure}[!htbp]
    \centering
    \includegraphics[width=1.0\columnwidth]{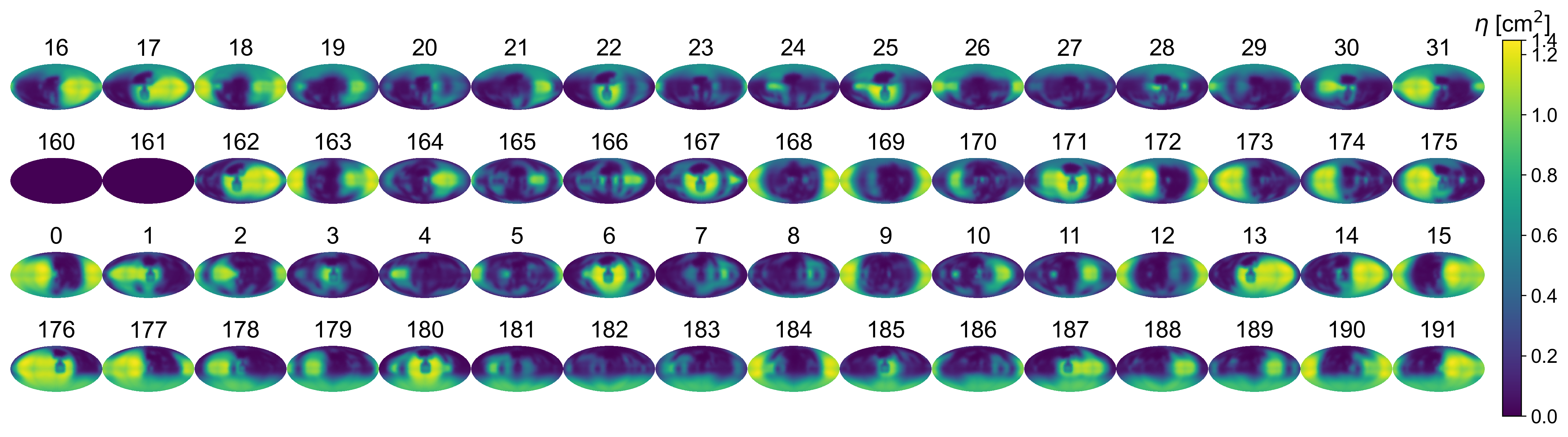}
    \caption{Simulated full-energy effective area $\eta(\theta, \phi)$ for $59$~keV gammas for each detector, indexed by their FPGA readout channel, shown in the Mollweide projection.
    In contrast to the simple numerical ordering of Fig.~\ref{fig:individual_fits_Cs137}, detectors are grouped by horizontal plane.
    Channels $160$ and $161$ have no detector module and thus have $\eta(\theta, \phi) \equiv 0$.
    }
    \label{fig:resp_59keV}
\end{figure}

\begin{figure}[!htbp]
    \centering
    \includegraphics[width=0.49\columnwidth]{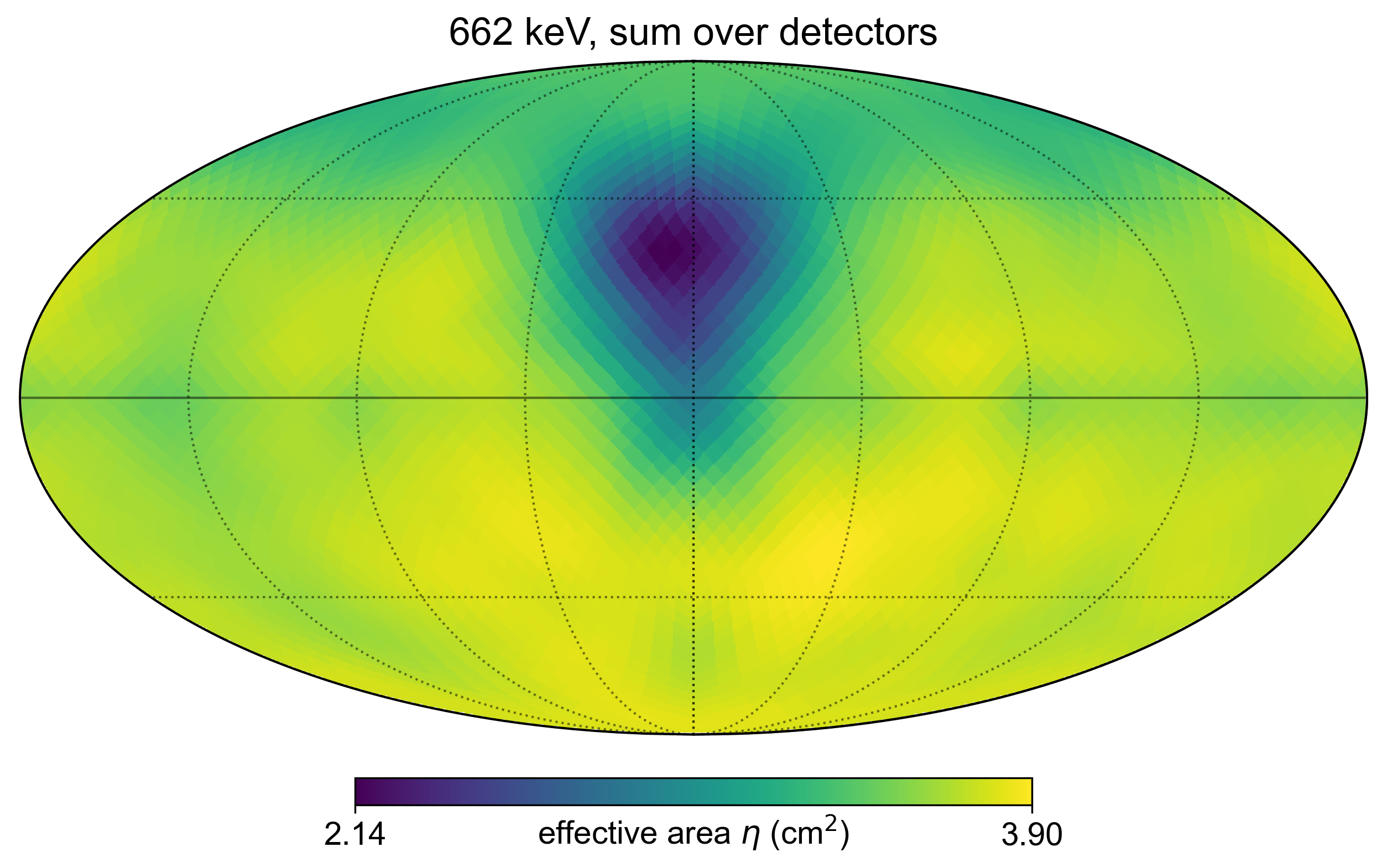}
    \includegraphics[width=0.49\columnwidth]{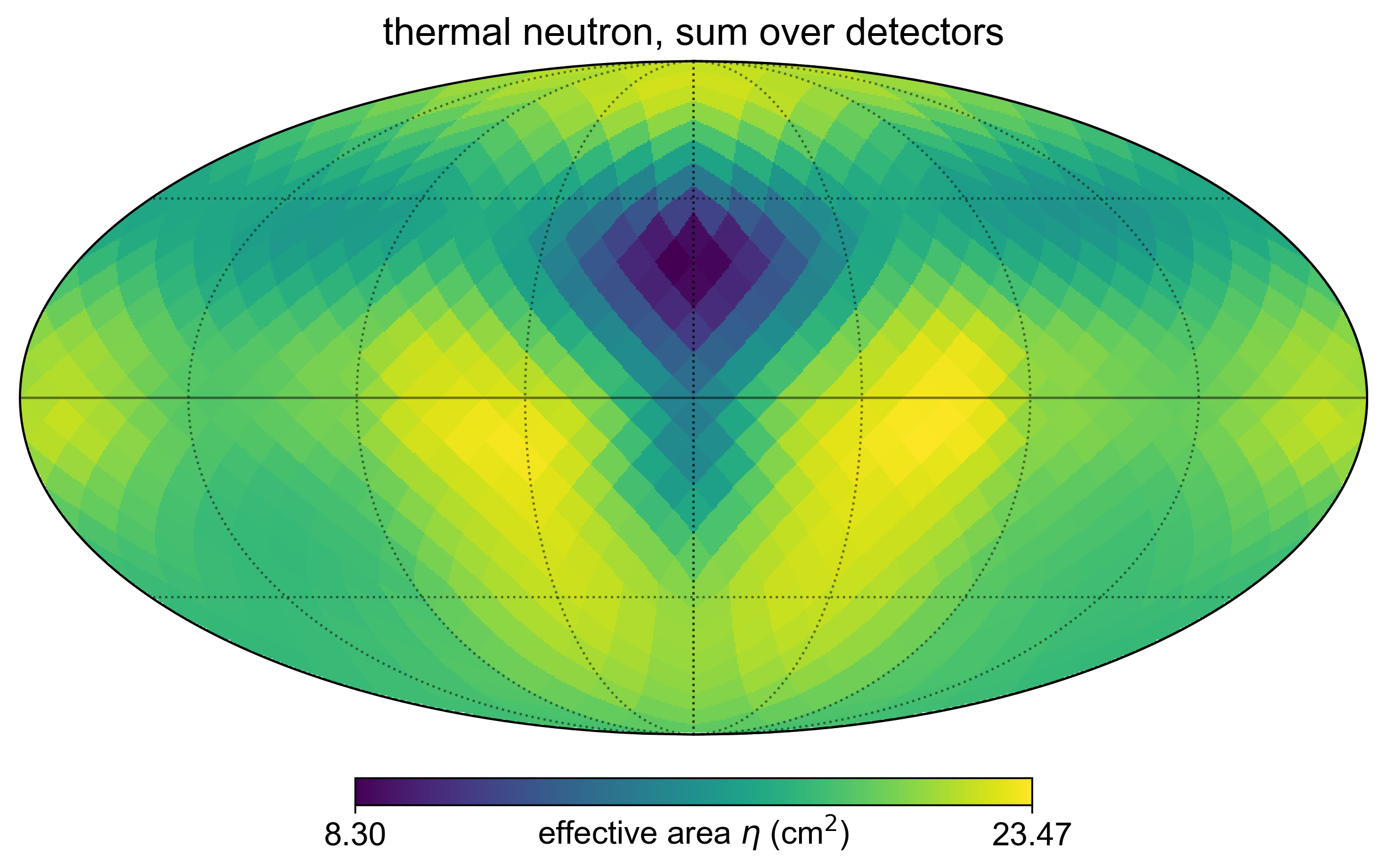}
    \caption{
        Simulated full-energy effective areas $\eta(\theta, \phi)$ in $4\pi$ for $662$~keV gammas (left) and thermal neutrons (right).
    }
    \label{fig:resp_sums}
\end{figure}

Fig.~\ref{fig:resp_59keV} shows the system's full-energy effective area for $59$~keV gammas for each individual detector module, while Fig.~\ref{fig:resp_sums} shows additional results for full-energy $662$~keV gammas and for thermal neutrons summed over all modules.
The darker, less efficient pixels correspond to attenuation from the LiDAR, LAMP battery, and, in the case of the individual detector plots, neighboring detector modules.
At $59$~keV the reduction in efficiency in these directions is about a factor of $6$ compared to the most efficient directions at the corners of the detector array, while at higher energies this ratio drops substantially due to higher photon penetrability.
The thermal neutron angular response tends to behave similarly to the $59$~keV gamma response due to the high neutron inelastic capture cross section of the CLLBC material.

\begin{figure}[!htbp]
    \centering
    \includegraphics[width=0.49\columnwidth]{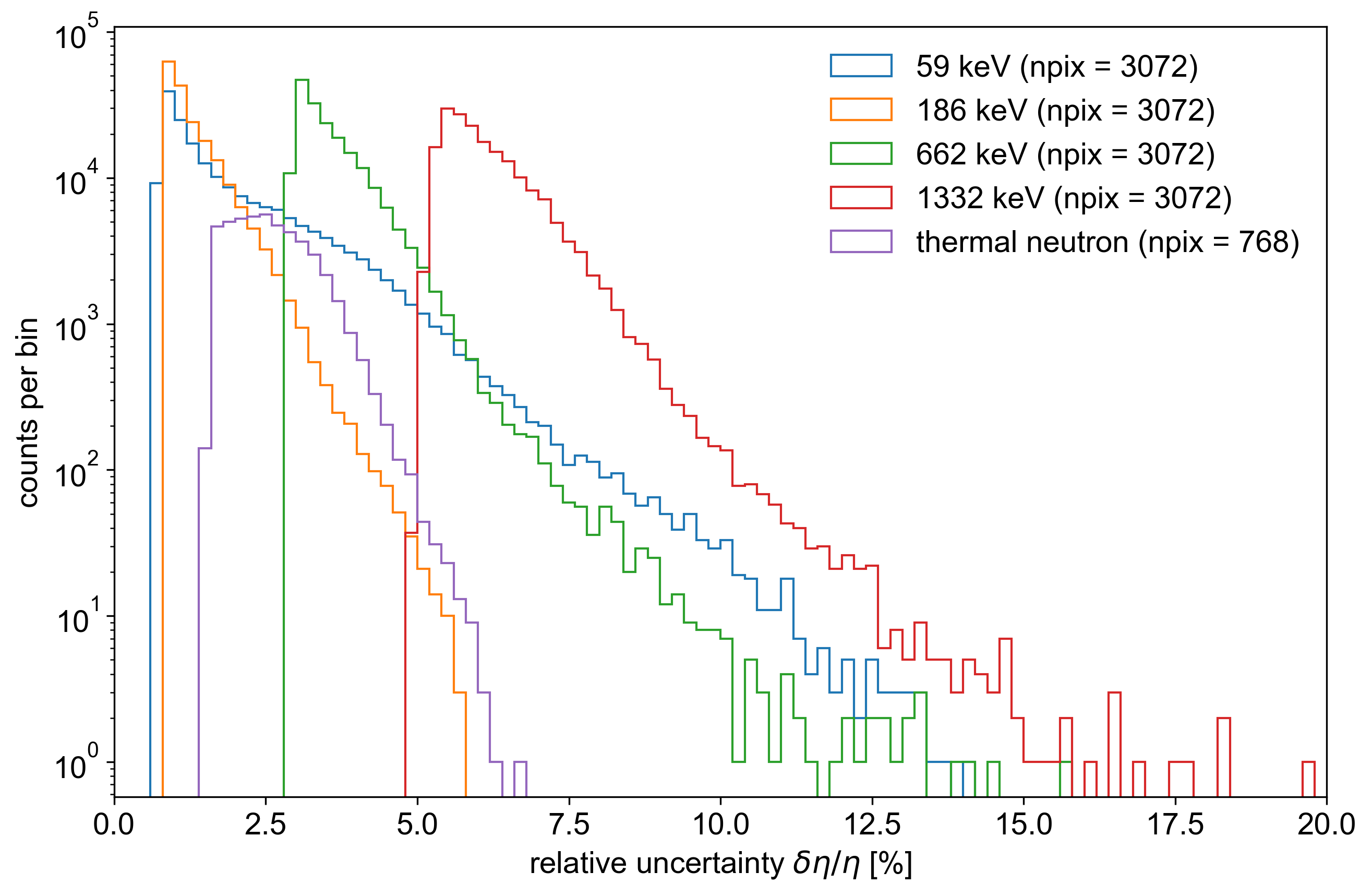}
    \includegraphics[width=0.49\columnwidth]{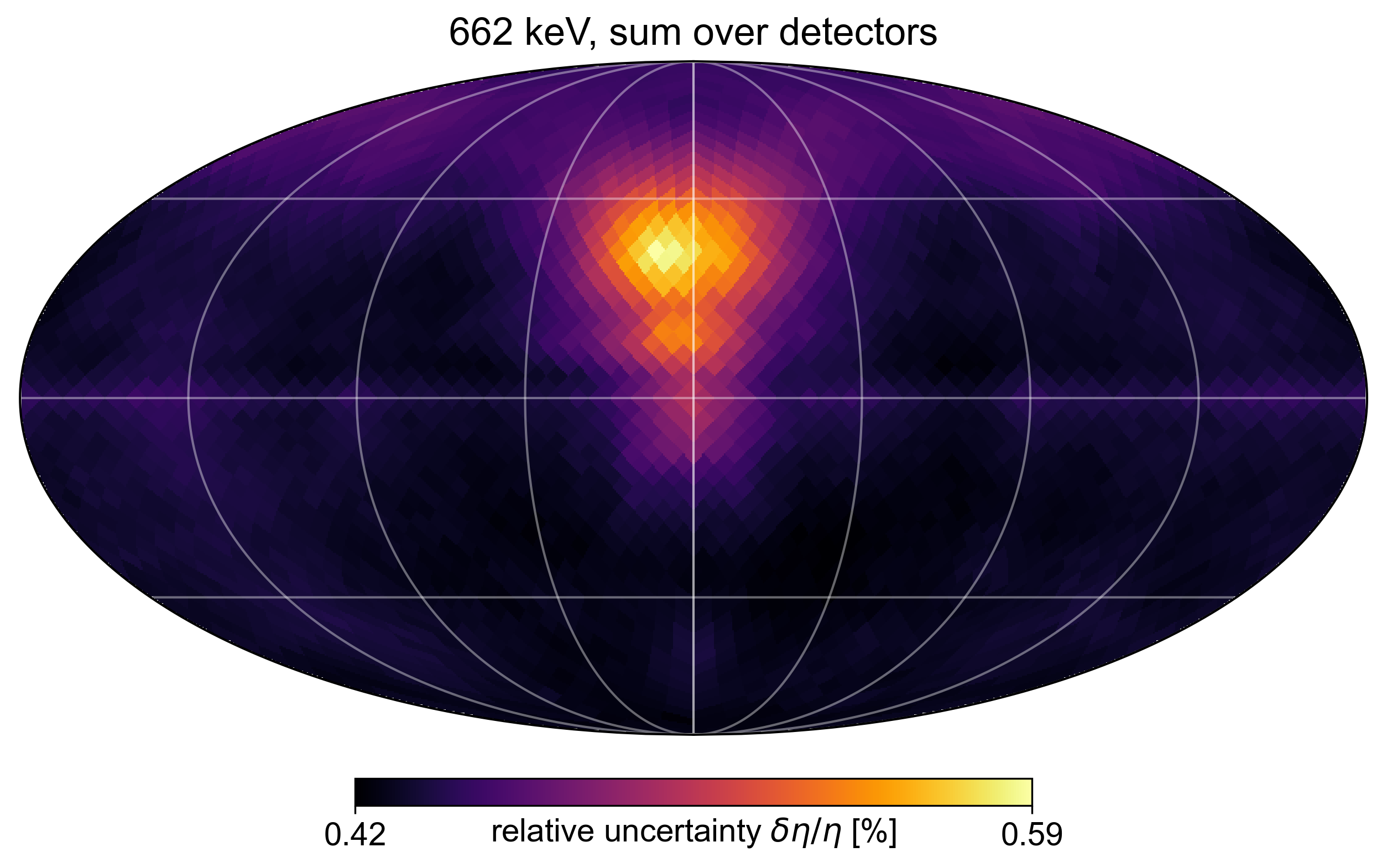}
    \caption{
        Left: simulated effective area relative statistical uncertainty $\delta \eta / \eta$ for all crystals and pixels, for selected primary particle types.
        Right: effective area relative statistical uncertainty summed over the array for $662$~keV gammas.
        Both plots use the unsmoothed response values.
    }
    \label{fig:stat_err}
\end{figure}

Fig.~\ref{fig:stat_err} shows the statistical errors in the Geant4 Monte Carlo response simulation for select primary particle types.
At the per-detector per-pixel level, most error distributions fall exponentially from a minimum value in the $1$--$5\%$ range, with only a small fraction of any pixels exceeding a $\delta \eta / \eta$ of $10\%$ for any primary particle type.
When summed over detectors, the per-pixel angular response uncertainties are reduced substantially to ${\sim}0.5\%$ (relative).

\begin{figure}[!htbp]
    \centering
    \includegraphics[width=0.99\columnwidth]{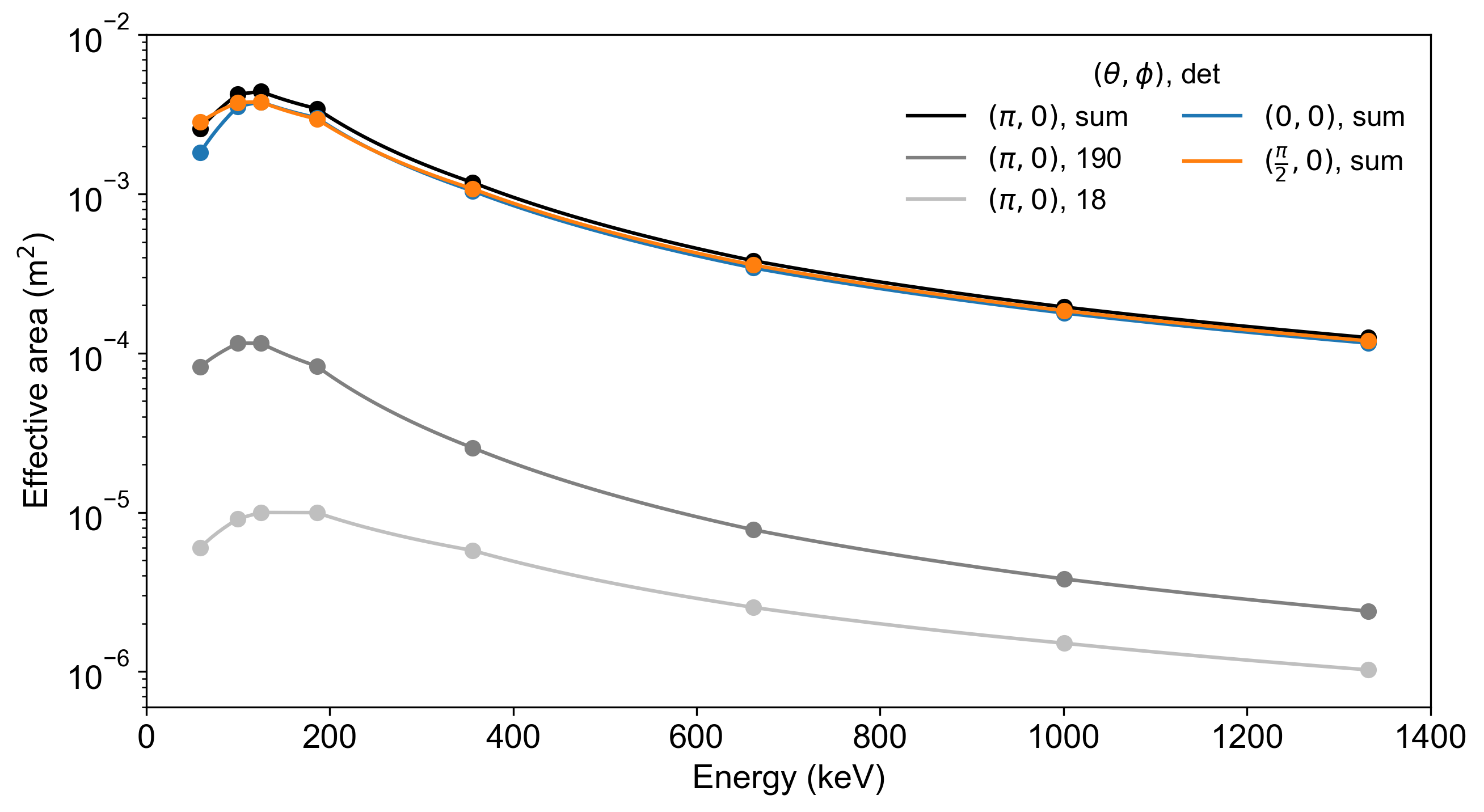}
    \caption{
        Simulated effective area $\eta$ as a function of photon energy $E$ at different angles and for different detectors.
        Circles show the simulated response values while the solid lines are a log-log interpolation between simulated values.
    }
    \label{fig:resp_vs_energy}
\end{figure}

Fig.~\ref{fig:resp_vs_energy} shows the simulated detector response $\eta$ instead as a function of energy for several different angles and detector modules.
In the downward-looking direction $(\theta, \phi) = (\pi, 0)$, detector~$18$ has one of the weakest responses in the entire array due to the presence of three crystals directly below it, including the unobstructed detector~$190$ on the bottom layer.
Summing over the detectors increases the response by a factor of $30$--$50$ over the $59$--$1332$~keV energy range, which is less than the number of detectors $J = 62$ due to attenuation from lower crystals.

\subsection{Validation measurement results}

Table~\ref{tab:gamma_validation} compares the imager's simulated effective areas for various gamma sources and directions to measured values from the set of validation experiments.
For \cs at $662$~keV, the simulated~$\eta$ are about $45\%$ higher than measured in the left, right, and back directions, rising to $96\%$ in the front direction where the LiDAR is located.
This discrepancy is slightly higher at around $50\%$ for the $1332$~keV line of \co, and lower at ${\sim}25\%$ for the $59$~keV line of \am.

\input{gamma_validation}

Table~\ref{tab:neutron_validation} compares the measured and simulated rates of thermal neutron inelastic events in the graphite measurements at various orientations.
In general, the simulated rates are ${\sim}70\%$ larger than the measured rates, a factor that is consistent across the front, back, left, right, and bottom irradiation directions.
Possible reasons for these discrepancies in both gamma and neutron responses will be discussed in Section~\ref{sec:discussion}.

\input{neutron_validation}

\subsection{Handheld simultaneous gamma and neutron localization}\label{sec:handheld_demos}

\begin{figure}[!htbp]
    \centering
    \includegraphics[width=0.95\columnwidth]{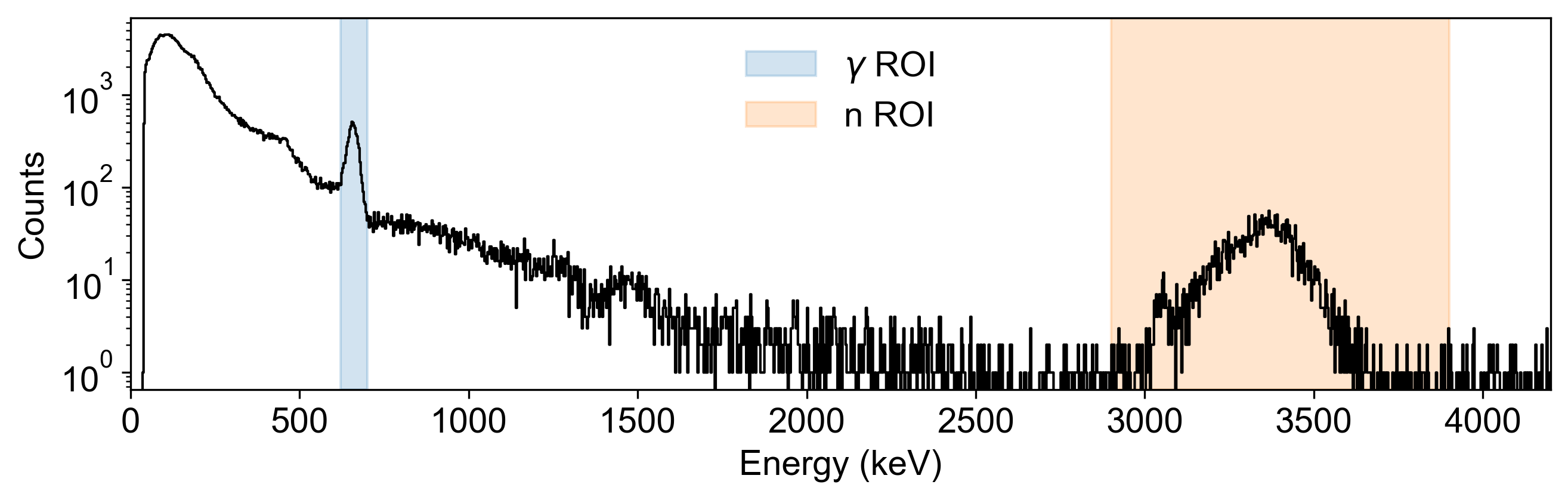}\\
    \includegraphics[width=1.00\columnwidth]{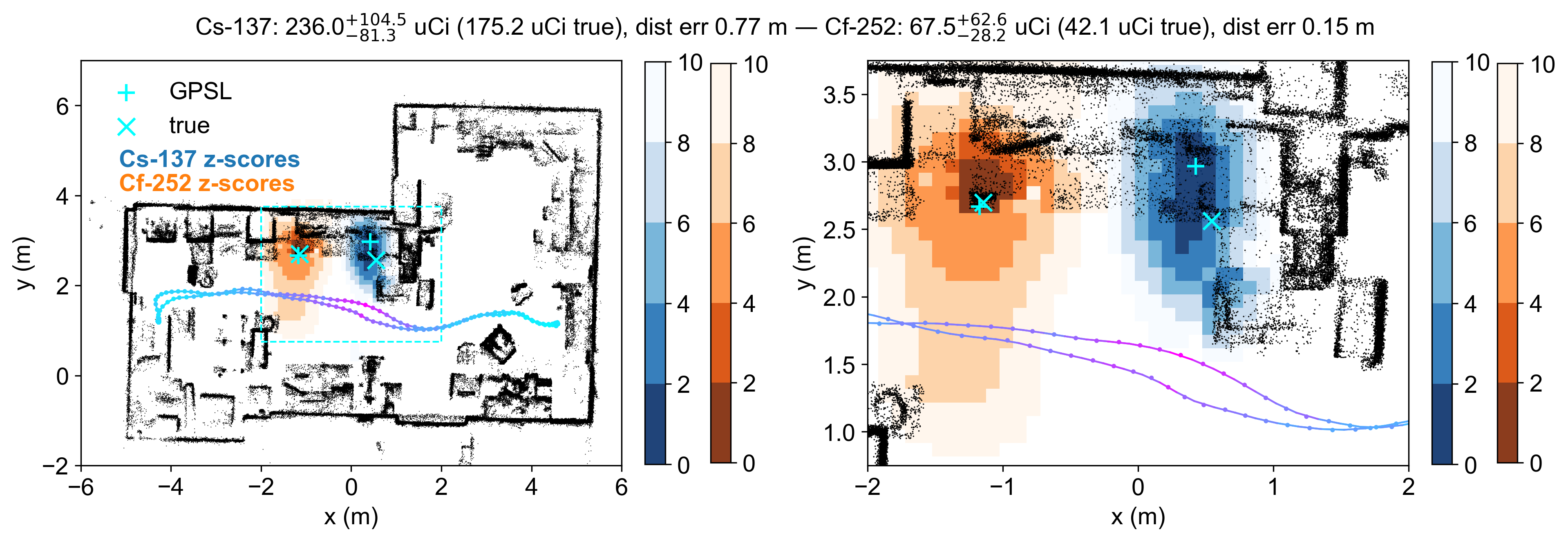}\\
    \includegraphics[width=1.00\columnwidth]{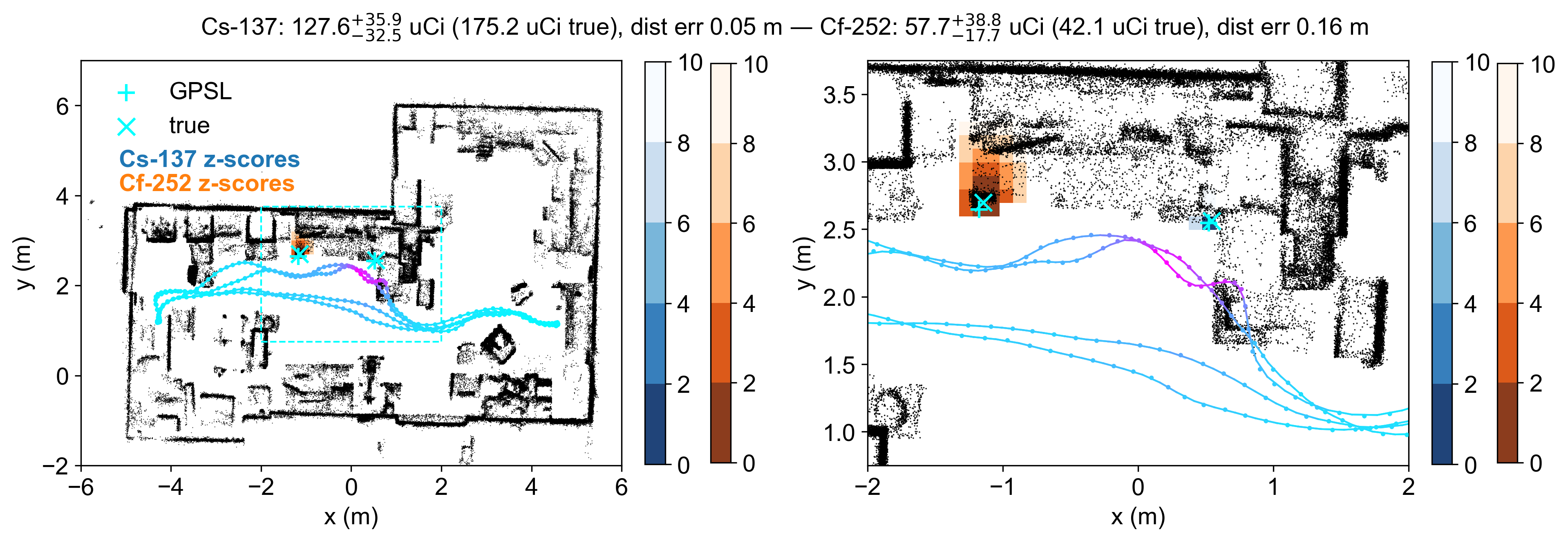}\\
    \includegraphics[width=0.95\columnwidth]{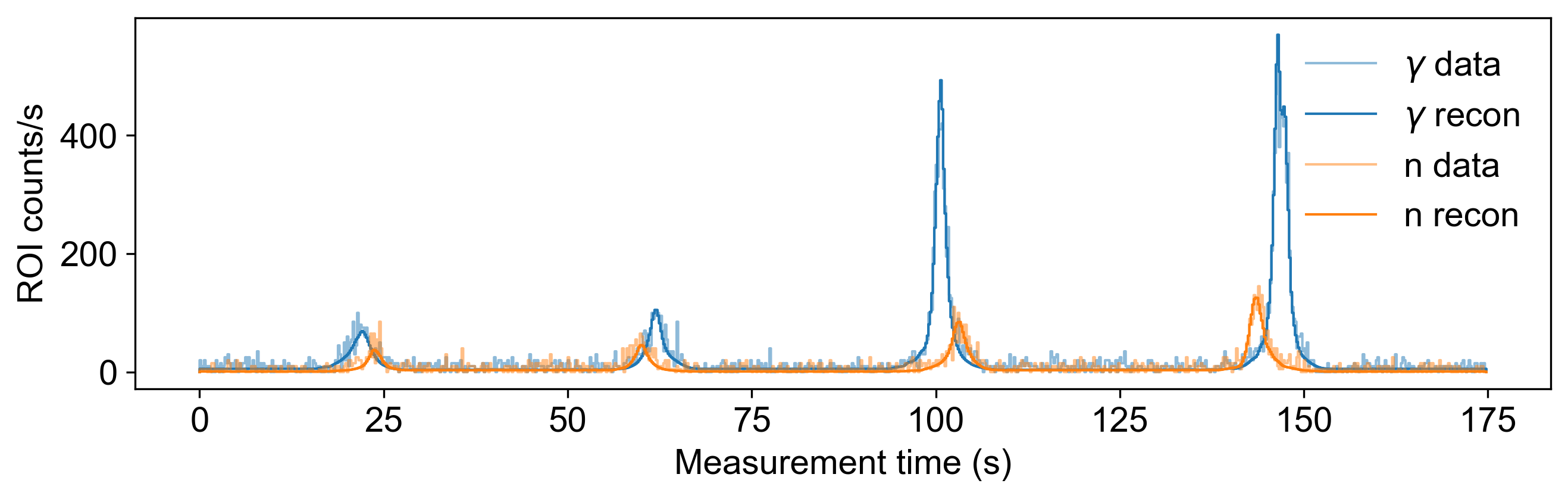}\\
    \caption{
        Handheld simultaneous measurement of a \cs source near $(x, y) = (0.5, 2.5)$~m and a \cf source near $(x, y) = (-1.0, 2.5)$~m.
        Top: energy spectrum, with the \cs and \cf ROIs denoted as blue and orange bands, respectively.
        Upper middle: top-down views of the LiDAR point cloud (black points), detector trajectory (colorized by gross counts), and combined individual GPSL reconstructions of the \cs and \cf sources using the first two pass-bys.
        GPSL $z$-scores are shown as the minimum along the $z$-axis projection.
        The most likely source position is shown in the cyan $+$ while the ground truth position is shown in the cyan $\times$.
        Activity uncertainties are given as $\pm 2\sigma$.
        Voxels are $10$~cm cubes.
        Lower middle: GPSL reconstructions using all four pass-bys.
        Bottom: measured and GPSL-reconstructed count rates in each ROI as a function of time.
    }
    \label{fig:ngv2_cs137_walkaround}
\end{figure}

In this section we demonstrate a handheld simultaneous \cs + \cf measurement.
A $175.2$~\uCi ($6.482$~MBq) \cs point source and a polyethylene-moderated $42.1$~\uCi ($1.56$~MBq) \cf source were placed ${\sim}1.5$~m apart on a benchtop in an indoor laboratory.
As shown in Fig.~\ref{fig:ngv2_cs137_walkaround}, the detector was hand-carried through the laboratory for ${\sim}180$~s, during which time it made four pass-bys of the sources.
In the first two passes, the detector was walked through the central lab corridor, remaining ${\sim}1$~m away from the sources.
GPSL reconstructions of both the \cs and \cf sources after two passes are shown in the upper middle of Fig.~\ref{fig:ngv2_cs137_walkaround}.
Even with the limited approach and left-right symmetry-breaking, the sources are localized to the correct side of the trajectory and in fact the correct corners of the benchtop, with position errors of $0.77$~m (\cs) and $0.15$~m (\cf).
The $0.77$~m \cs error in particular is largely due to a $0.65$~m error in the $z$ direction, resulting from the limited up-down symmetry-breaking from the nearly-level detector trajectory.
In the last two passes, the detector more closely surveyed the benchtop, approaching to within $0.5$~m of the sources.
GPSL reconstructions using all four passes are shown in the lower middle of Fig.~\ref{fig:ngv2_cs137_walkaround}.
The GPSL likelihood contours are noticeably reduced compared to the two-pass reconstructions, as are the activity uncertainty estimates, and the position errors drop to $0.05$~m (\cs) and $0.16$~m (\cf).
The \cs activity estimate is biased $27\%$ low and the ground truth activity is slightly outside $2\sigma$ confidence interval.
Conversely the \cf activity estimate is $37\%$ higher than the true value, but consistent within $2\sigma$, even without taking into account the moderator geometry, which might be unknown in a real search scenario.

Although the reconstructions here were run offline after the measurements, this demonstration emulates a scenario where the reconstructions can guide the detector operator in near-real-time during the measurement.
The GPSL reconstruction after just the first pass ran in ${<}2$~s and is sufficiently informative (see also Section~\ref{sec:response_degradation}) to direct the operator to the correct side of the laboratory for further survey.
Adding additional passes increased the accuracy of the position localization, but no reconstruction time exceeded $4$~s.

\subsection{Vehicle-borne demonstrations}\label{sec:vehicle_demos}
\begin{figure}[!htbp]
    \centering
    \includegraphics[width=1.0\columnwidth]{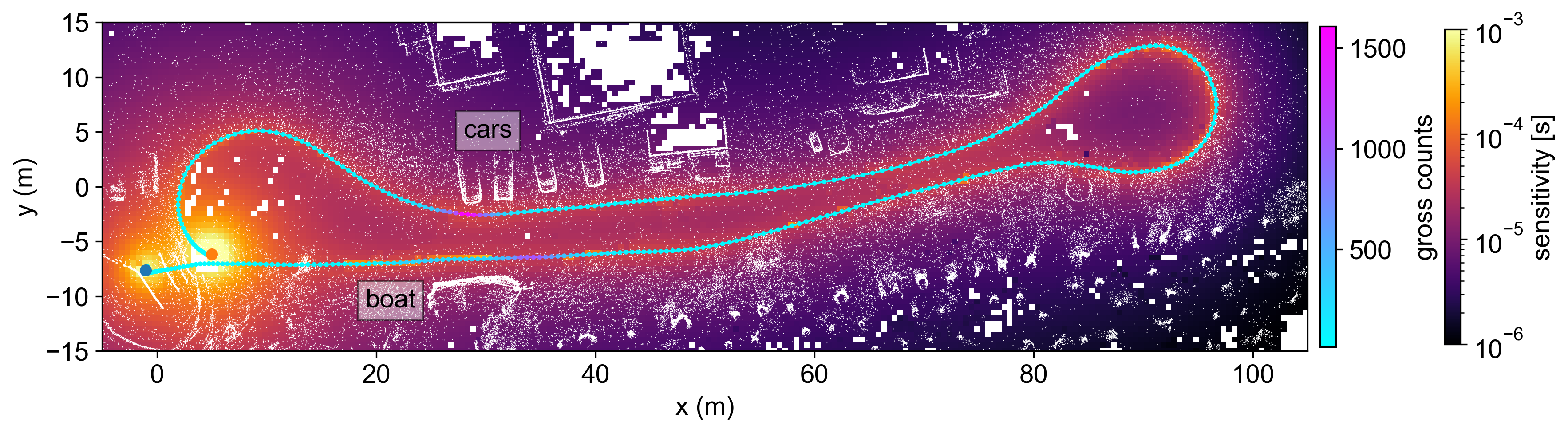}\\
    \includegraphics[width=1.0\columnwidth]{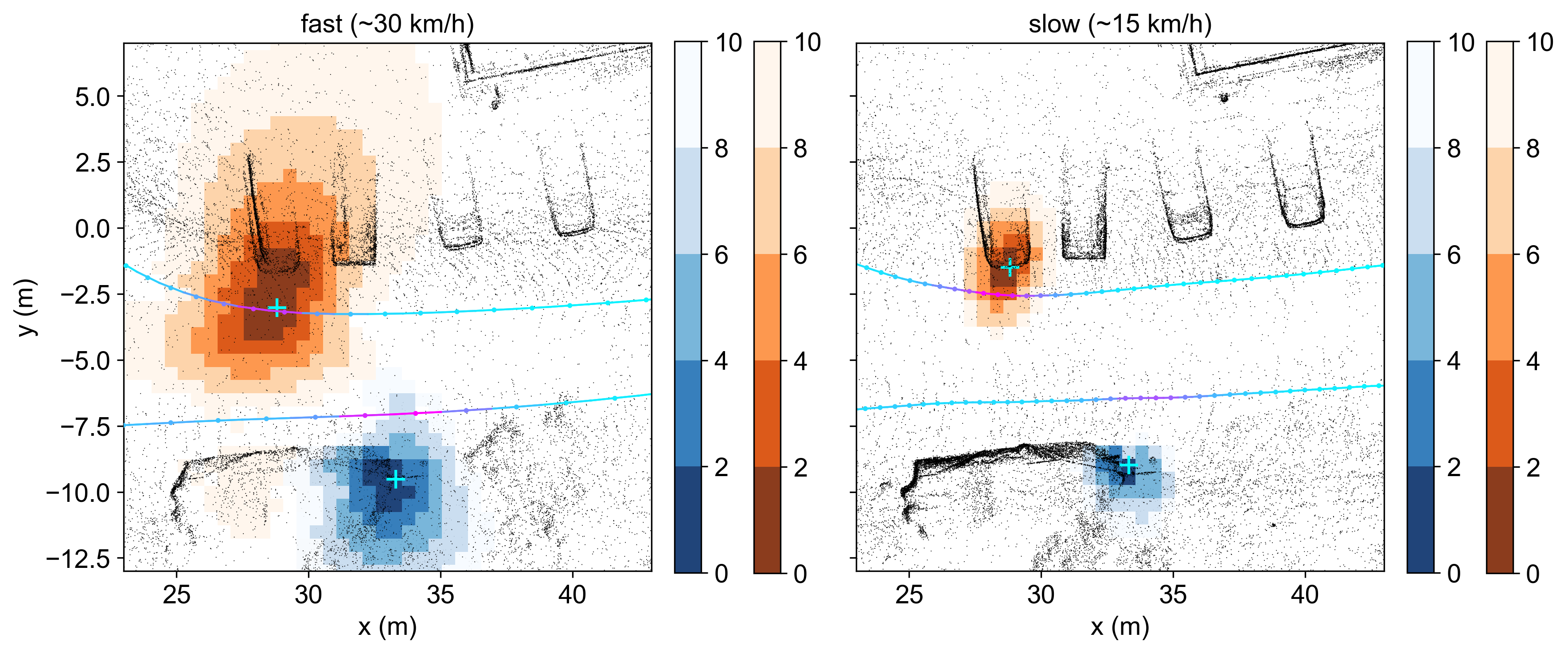}\\
    \includegraphics[width=1.0\columnwidth]{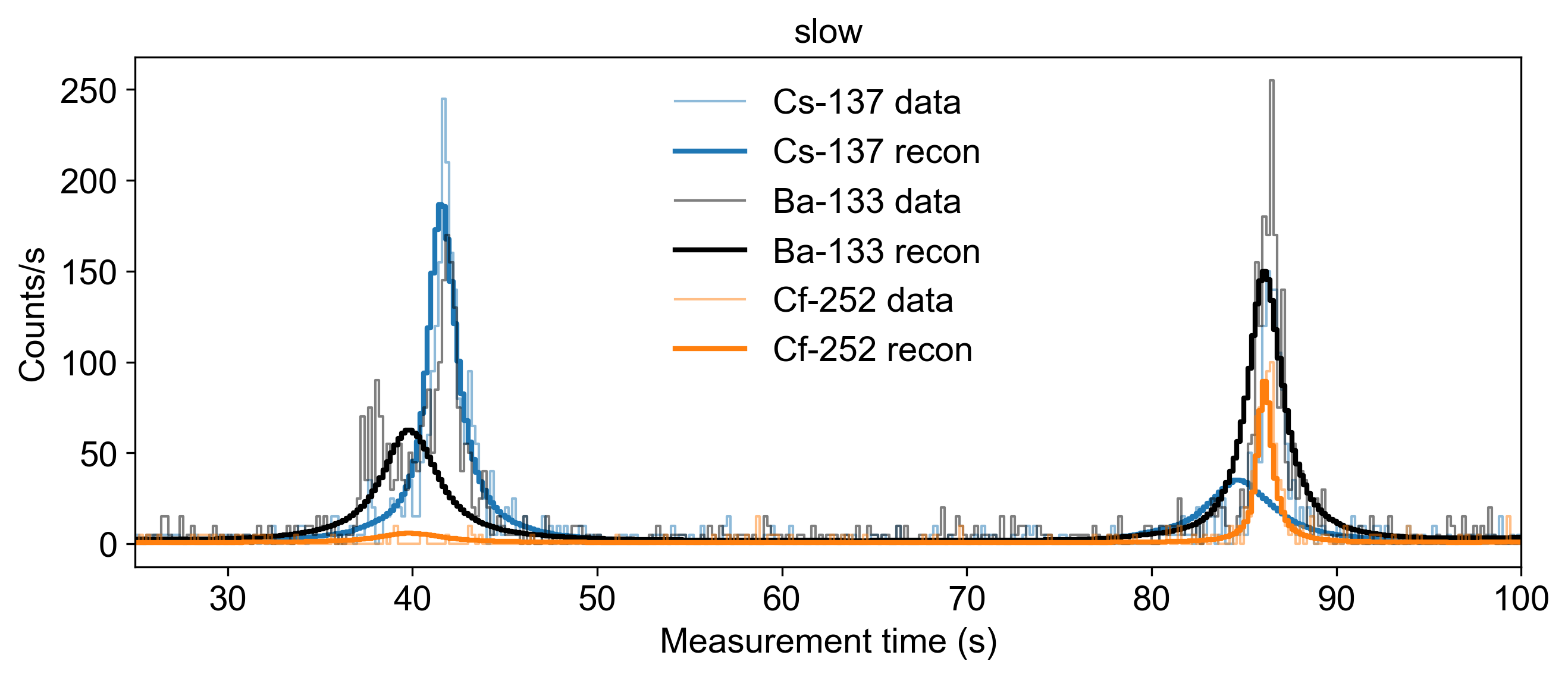}
    \caption{
        Gator radiation surveys of a stretch of road using the imager.
        Top: zoomed-out view of the slow measurement, showing the LiDAR point cloud (white points), the sensitivity map $\varsigma$ (max reduction in the $xy$ plane), and spikes in gross counts near the source positions.
        Middle: GPSL reconstructions of \cs (blue) and \cf (orange) in the fast (left, ${\sim} 30$~km/h) and slow (right, ${\sim} 15$~km/h) drive-by.
        The highest-likelihood voxel in each reconstruction is indicated with a cyan $+$ marker.
        For visual clarity, the \ba reconstruction contours are not shown.
        Voxels are $50$~cm cubes.
        Both measurements were transformed to the same coordinate frame using the Iterative Closest Point (ICP) method.
        Bottom: measured and GPSL-reconstructed count rates in the three ROIs during the slow measurement.
    }
    \label{fig:gator_drive-bys}
\end{figure}

In addition to handheld surveys, vehicle-borne surveys are a very attractive mode of operation for in-field detection and emergency response, enabling fast searches of wide areas and reduced dose to personnel.
In this series of measurements, the imager was driven along a stretch of road at the Richmond Field Station in Richmond, CA on a Gator vehicle at nominal top speeds of $15$, $20$, and $30$~km/h.
Various sources were hidden in vehicles on either side of the road---a plutonium surrogate source (a combination of \cf, \cs, \ba, and \am) as well as an additional shielded \cs source were placed in the back of a car, while additional \ba and \cs sources were placed in a boat on the opposite side of the road.
For the purposes of this demonstration, although the precise source positions, activities, and shielding configurations were not recorded as they were in Section~\ref{sec:handheld_demos}, it is still possible to localize sources and discuss qualitative trends at different speeds.
Fig.~\ref{fig:gator_drive-bys} shows GPSL reconstructions of the \cs and \cf source positions in the fast and slow measurements.
The \cf is correctly localized to the trunk of the correct vehicle in the slow measurement, while the likelihood contours in the fast measurement are large enough that the left or right side of the trajectory cannot be unambiguously determined.
The \cs reconstruction performs similarly, with the likelihood contours contracting at slower speeds due to more time spent near the source, though the reconstruction only picks out one of the two \cs sources present.
Similar general trends are observed for the \ba reconstructions (not shown), although in this case the presence of two sources results in the \ba reconstruction being placed in the middle of the road between the two sources rather than picking out one or the other.
In this scenario, the single-point-source model inherent to GPSL is an inadequate model for the measurement.
This limitation is further demonstrated by comparing the reconstructed \ba count rates in Fig.~\ref{fig:gator_drive-bys} to the measured data.
Multiple-point-source reconstructions with the imager will be demonstrated in Section~\ref{sec:apsl}, and, although not an exact analogue to this two-source issue, the question of left-right symmetry breaking will be more thoroughly explored in Section~\ref{sec:response_degradation}.

\begin{figure}[!htbp]
    \centering
    \includegraphics[width=1.00\columnwidth]{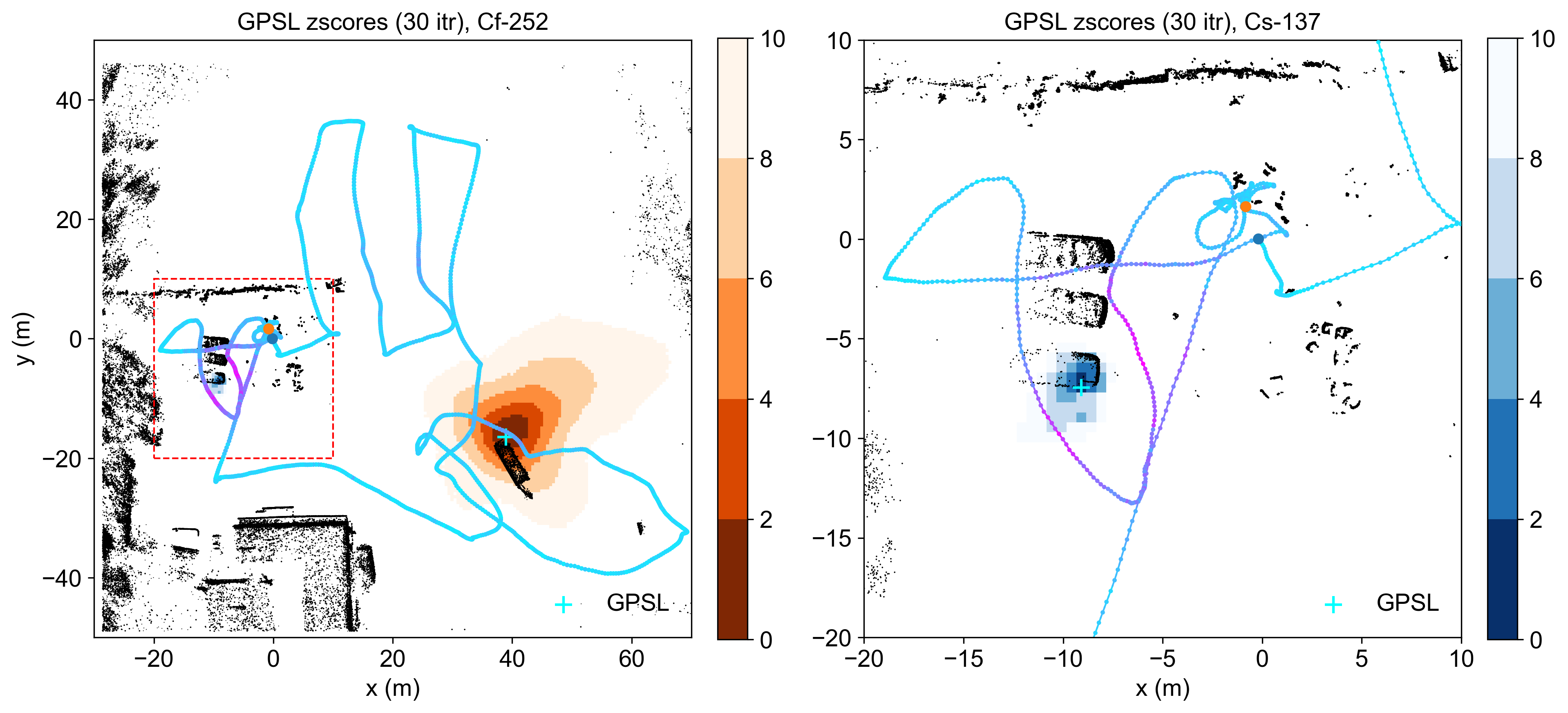}
    \caption{
        UAS radiation survey of a field using the imager.
        Left: GPSL reconstruction of the \cf component of a plutonium surrogate source in a van.
        Right: zoom in on the dashed red square of the left panel, showing the GPSL reconstruction of a shielded \cs source in a car.
        The blue and orange points near $(x, y) = (0, 0)$ indicate the start and stop of the trajectory, respectively.
        In both panels, only point cloud data with $z>0$ is plotted for visual clarity, though no points are excluded from the reconstruction analysis.
        The cyan $+$ markers denote the highest likelihood voxel in each reconstruction.
        Voxels are $50$~cm cubes.
    }
    \label{fig:first_flight}
\end{figure}

In another vehicle-borne demonstration, a human pilot flew the imager on a small UAS over a field where a shielded \cs source was hidden in one car and the plutonium surrogate source was hidden in a van ${\sim}50$~m away.
The system flew for $400$~s typically between $4$--$5$~m above ground level (AGL), where it both surveyed the nearby vehicles and performed a short raster over the empty field.
GPSL reconstructions were then run on the \cs and thermal neutron datasets collected.
Because the source shielding configurations are again unknown (and enhanced by the vehicles themselves), we do not attempt to quantify the activity of the sources here, but as shown in Fig.~\ref{fig:first_flight}, GPSL indeed localized the sources to the correct vehicles.
In particular, the GPSL \cs reconstruction places essentially zero likelihood on the two cars parked next to the source-bearing car, which provides sufficient attribution to direct further search activity.
Similarly, the GPSL neutron reconstruction correctly finds the neutron source in the van, and in fact correctly localizes it to the back of the vehicle.
The likelihood contours are however substantially larger than those of the \cs reconstruction, likely due the lower statistics in the thermal neutron signal as well as thermalization in the ground and vehicle's gas tank.

\subsection{Response degradation analysis}\label{sec:response_degradation}

In this section, we demonstrate the value of the full omnidirectional anisotropic multi-crystal response in breaking the source reconstruction degeneracies discussed in Section~\ref{sec:math}.
To do so, we construct an imaging scenario in which the detector is constrained to three forward-and-back passes along a level, straight line of about $5$~m in length and where a single, relatively strong ($178.1$~\uCi) \cs point source is present ${\sim}1.5$~m to the side approximately midway along the detector path.
Such a setup could represent, say, a detector placed on a fixed translation stage for remote safeguards of nuclear facilities, or a detector operating in a vehicle travelling along a straight stretch of road (similar to Fig.~\ref{fig:gator_drive-bys}).
In this scenario, a detector response that cannot break spatial degeneracies will result in a reconstruction that can determine the detector's location and distance of closest approach to the source, but cannot determine the source's position along the ring of points at that distance.
For instance, in Fig.~13 of Ref.~\cite{bukartas2022accuracy}, a \co source could be detected with high confidence after a single vehicle pass-by, but could not be reliably localized to the correct side of the road due to the lack of position sensitivity of the single high-purity germanium (HPGe) detector used.
In addition, we will compare localization performance both with and without the LiDAR occupancy map, which is highly relevant to measurements in facilities where such LiDAR measurements are not permitted.

Therefore, to demonstrate the imager's ability to break spatial degeneracies, we compare its GPSL reconstruction in this linearly-constrained scenario to GPSL reconstructions where we have degraded its response in order to emulate less-capable detector systems.
In particular, the degradations considered are
\begin{enumerate}
    \item summing the response over detector elements (``monolithic''),
    \item summing the response over detector elements and forcing the response to be isotropic,
    \item removing the LiDAR occupancy cut, i.e., reconstructing on \textit{all} voxels rather than occupied voxels,
\end{enumerate}
where either methods 1 or 2 may also be combined with 3.
Figs.~\ref{fig:degraded_walkbys} and \ref{fig:degraded_walkbys_lidar_cut} show the GPSL reconstructions at three decreasing levels of degradation (2+3, 1+3, 2) and no degradation.
In the first case, the detector response is monolithic and isotropic (and uses no LiDAR occupancy cut) and thus cannot break the ring of degeneracy at the point of closest approach, as shown by the ring of highly-probable ($<2\sigma$) potential source voxels.
There is a modest bias in the ring contours towards the correct side of the trajectory (perhaps due to the up to $10$~cm deviations around the idealized straight path inducing small, unintended variations in~$r^2$), and the reconstructed distance of closest approach of $1.29$~m is close to the true value of $1.23$~m, though the reconstructed $z$ position is off by $0.85$~m.
Using the summed monolithic response alone (but still all voxels) slightly reduces the size of the ${\leq}2\sigma$ contour ($n_{2\sigma}$ in Fig.~\ref{fig:degraded_walkbys}) compared to the monolithic+isotropic response, but still cannot break the angular degeneracy and therefore does not substantially change reconstruction performance.
Conversely, using the full multi-crystal detector response (without the LiDAR occupancy cut) significantly reduces the number of likely voxels, improves the reconstructed location error from $0.89$~m to $0.21$~m, and reduces both the activity error and confidence interval, but still has a modest spread in likely voxels in the $z$ dimension due to the lack of $z$ motion of the detector.
Finally, using the full response plus the LiDAR occupancy cut (Fig.~\ref{fig:degraded_walkbys_lidar_cut}) returns the best-performing reconstruction, with a position error of $0.15$~m, the smallest likelihood contours, and the smallest activity confidence interval.
Therefore, in these linearly-constrained, degenerate scenarios, the largest performance gains are found in using the multi-crystal (i.e., position-sensitive) response, though typically at the expense of an order of magnitude increase in reconstruction time.

\begin{figure*}[!htbp]
    \centering
    \includegraphics[width=1.00\columnwidth]{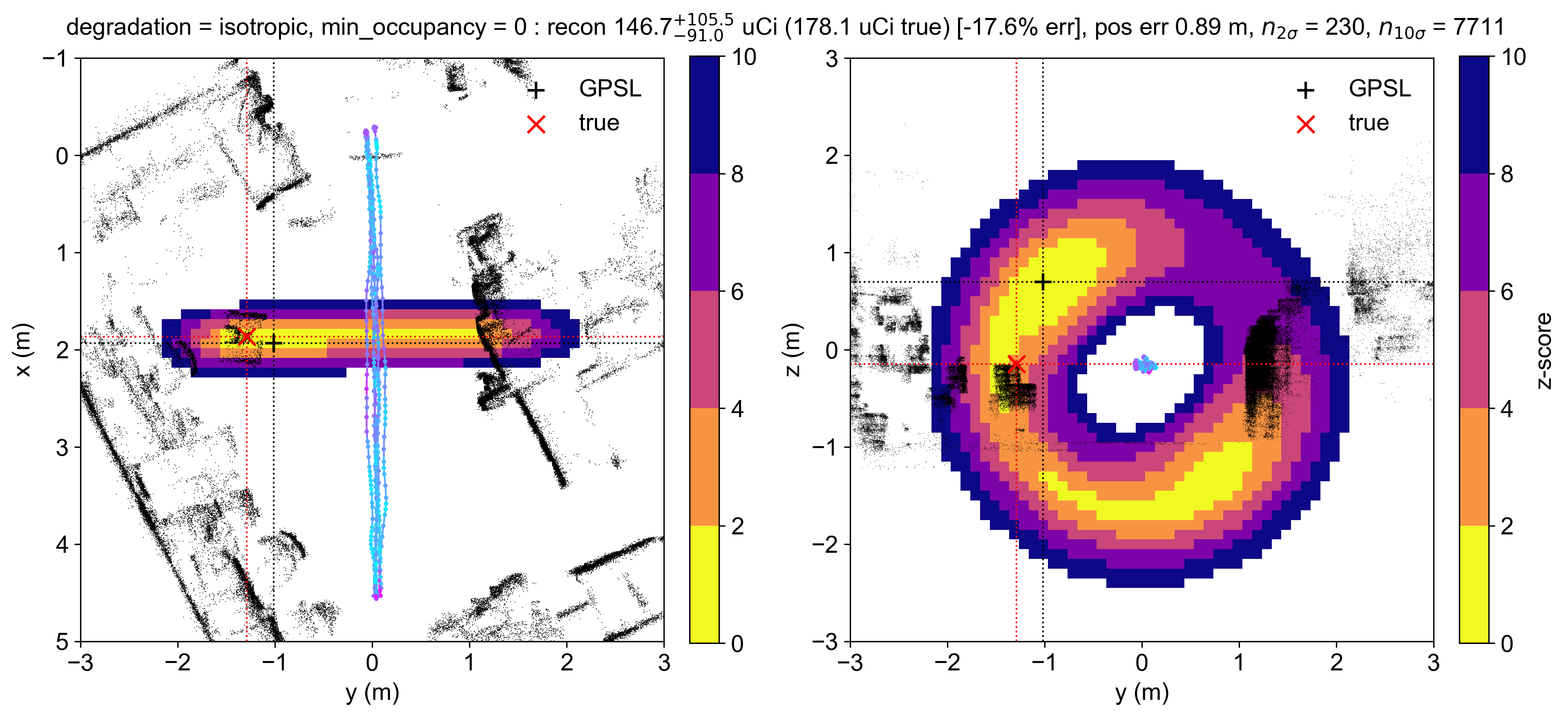}
    \includegraphics[width=1.00\columnwidth]{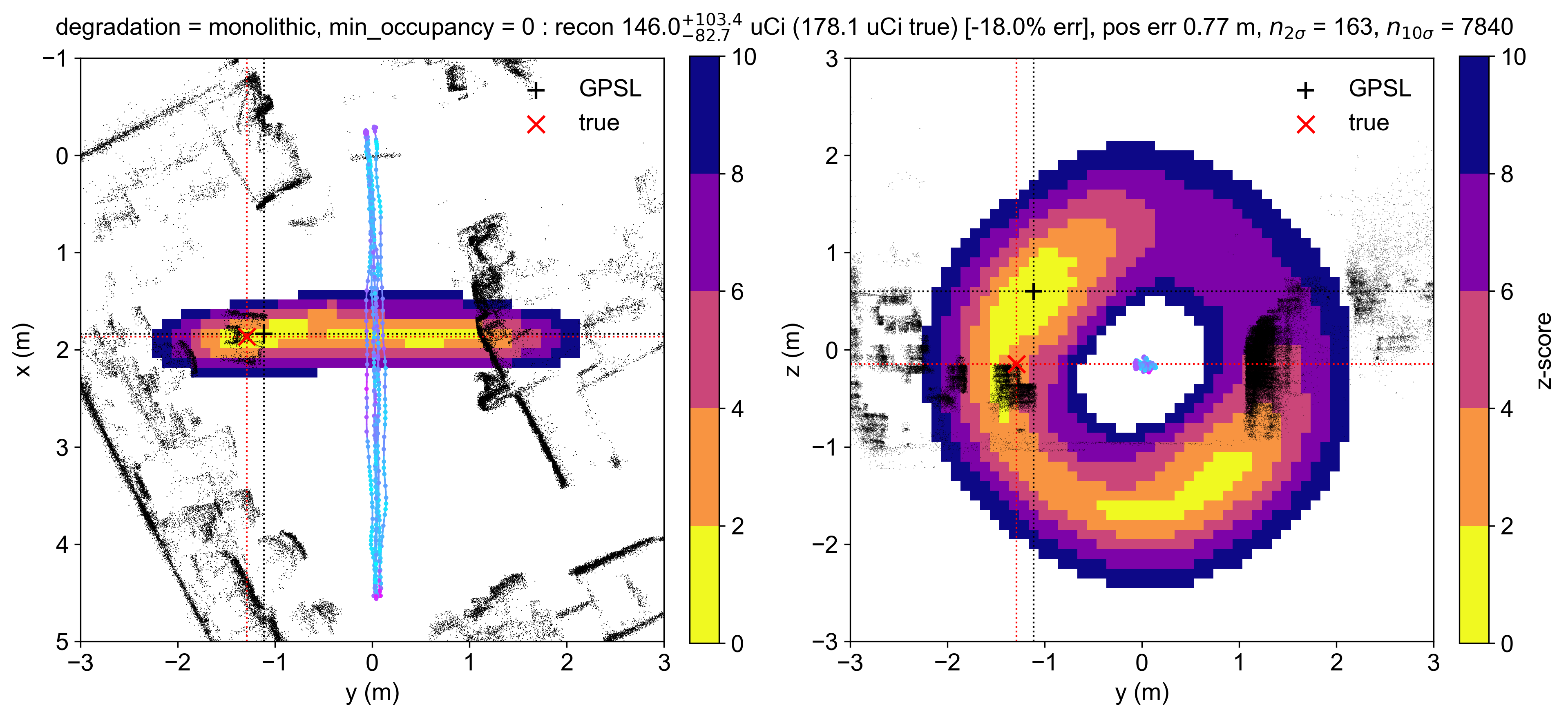}
    \includegraphics[width=1.00\columnwidth]{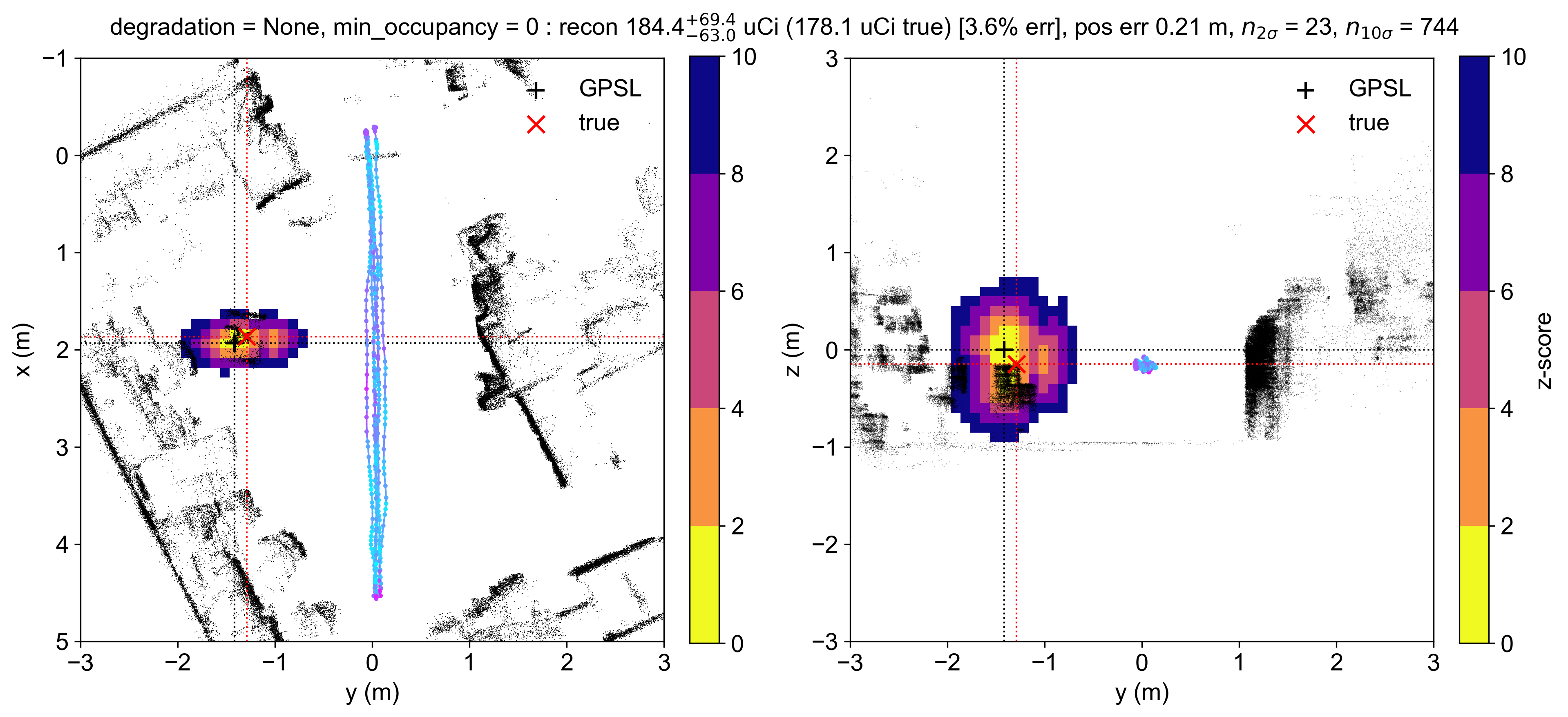}
    \caption{
        GPSL reconstructions at three decreasing levels of response degradation, with color maps denoting the minimum $z$-score at each pixel under a $yx$ or $yz$ projection (left and right columns, respectively).
        Each reconstruction was run for $100$~iterations with $10$~cm cubic voxels.
        Top: isotropic detector response, no LiDAR occupancy cut.
        Middle: monolithic detector, no LiDAR occupancy cut.
        Bottom: full detector response, no LiDAR occupancy cut.
        Cuts of $z \in [-0.9, 0.9]$~m and $x \in [1.5, 2.5]$~m are applied to the $yx$ and $yz$ point cloud plots, respectively, for visual clarity, but not are not applied in the analysis.
        Activity confidence intervals are $\pm 2\sigma$.
    }
    \label{fig:degraded_walkbys}
\end{figure*}

\begin{figure*}[!htbp]
    \centering
    \includegraphics[width=1.00\columnwidth]{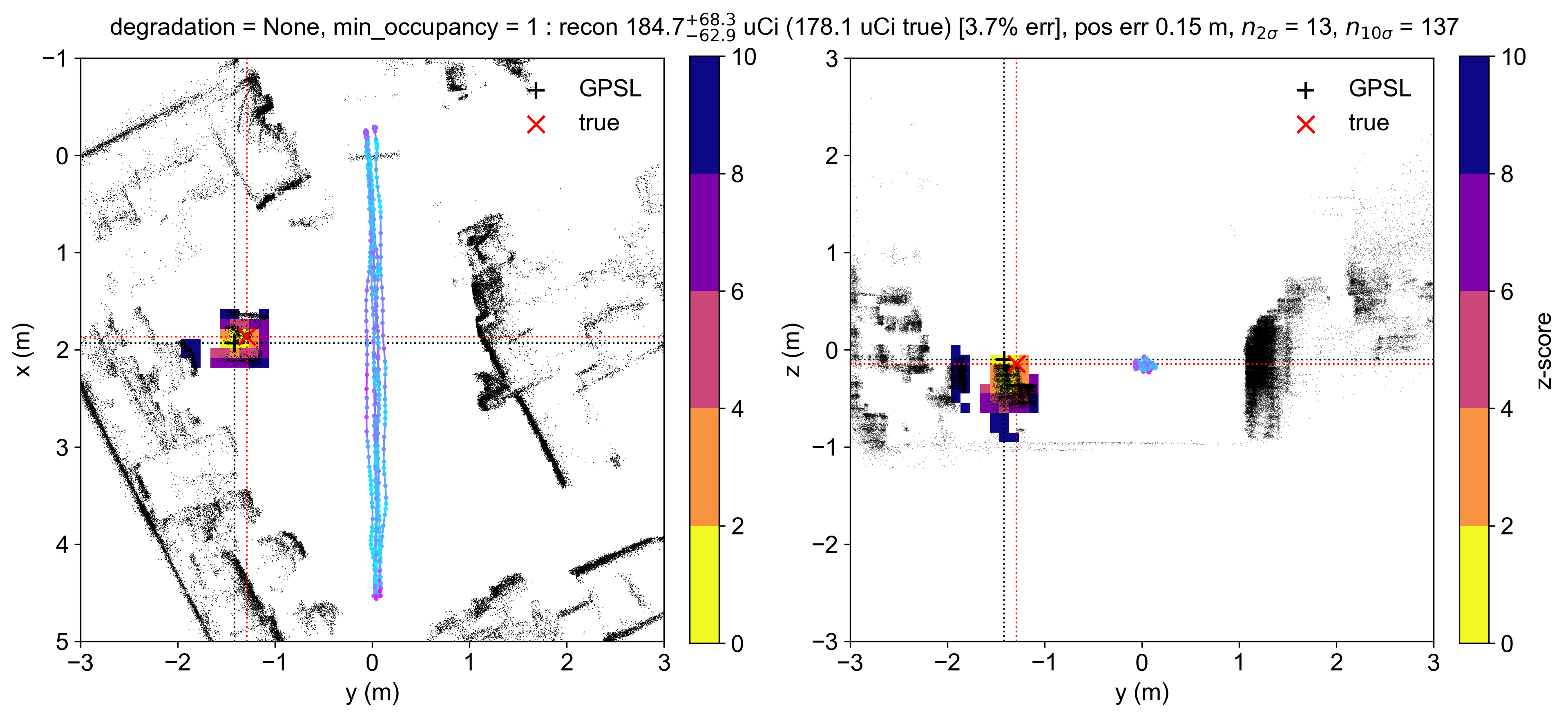}
    \caption{Continuation of Fig.~\ref{fig:degraded_walkbys}: full detector response, with LiDAR occupancy cut.}
    \label{fig:degraded_walkbys_lidar_cut}
\end{figure*}

We note here that the point cloud in Fig.~\ref{fig:degraded_walkbys} was generated by performing a dedicated free-moving LiDAR scan of the entire laboratory prior to placing the imager on a cart for the constrained measurements.
However, the radiation and trajectory data during the free-moving scan was excluded from the GPSL reconstruction in order to focus on the constrained pass-bys.
The measurement time after this cut was~$80$~s.
Finally, we note that the pass-bys shown here kept the detector facing the same way all the time.
In a similar set of measurements where the detector was rotated $180^\circ$ after each pass-by, the overall performance trends were similar to the unrotated cases shown in Fig.~\ref{fig:degraded_walkbys}.

\subsection{Static singles imaging}\label{sec:static_singles}

In most of the previous demonstration measurements, we have depended primarily on the system's (free-) moving capability to break the $r^2$ degeneracies inherent in Eq.~\ref{eq:lambda}.
Here we show that, contrary to claims about SDF in the literature (e.g., \cite{zhao2022two}), SDF strongly leverages but does not \textit{necessarily} rely on movement through the scene.
The full-fidelity omnidirectional multi-crystal response coupled with GPSL is in fact powerful enough to enable \textit{static} singles imaging, i.e., the imaging of a point source from a \textit{stationary} detector using only singles-mode (non-Compton) data, because in GPSL, there is no conceptual difference between multiple static detector modules at different locations and one module making successive measurements at those locations.
As such, the imager in static singles mode can accurately determine the direction to a point source via the usual 3D GPSL reconstruction.
With enough statistics, the small but nonzero separation between detector modules (and their active masking of each other) should even enable the estimation of the source's distance (and thus activity).
In our case, however, the GPSL distance estimates proved unreliable (possibly due to the far-field response approximation), and so in this section we focus primarily reconstructing the angle to the source.
We do however show that applying the LiDAR occupancy cut can accurately determine the source distance along the reconstructed angle.

\begin{figure}[!htbp]
    \centering
    \includegraphics[width=1.0\columnwidth]{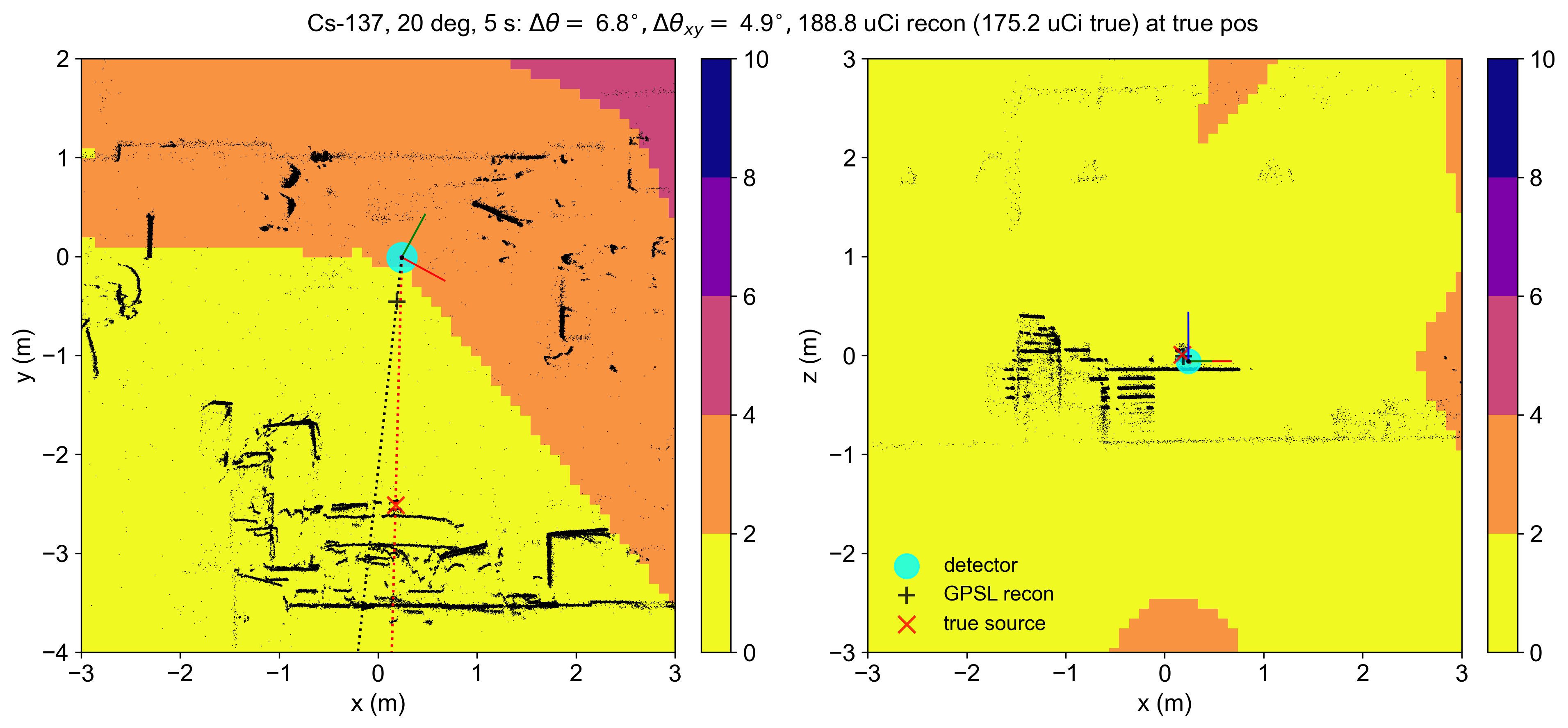}\\
    \includegraphics[width=1.0\columnwidth]{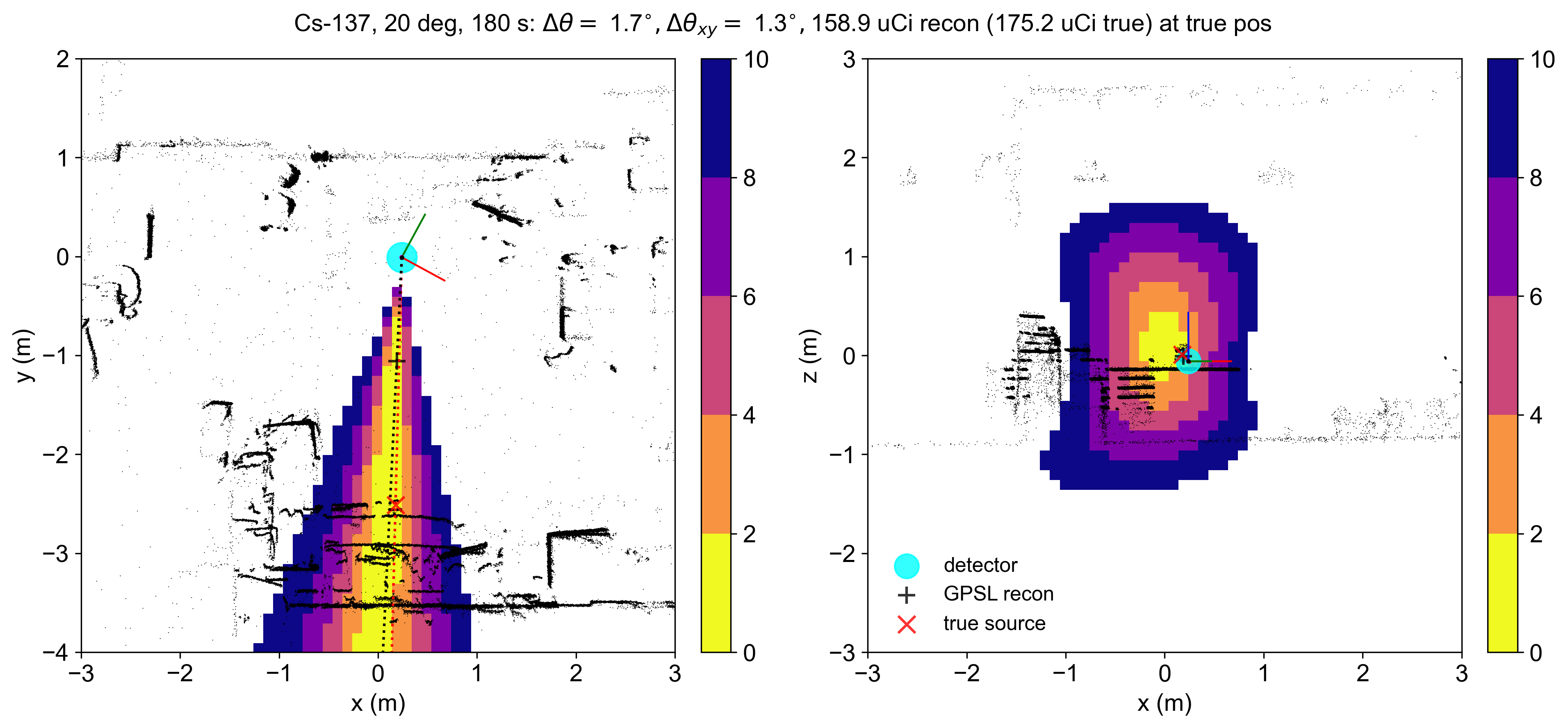}
    \caption{
        Static singles imaging of a \cs point source with the imager rotated a nominal $20^\circ$ angle to the source.
        The color bars give the GPSL $z$-scores under minimum projections in the $xy$ (left) and $xz$ (right) planes, and the top and bottom rows show GPSL results after $5$~s and $180$~s dwell times respectively.
        The red, green, and blue lines extending from the detector marker (cyan circle near $(0, 0, 0)$) denote the $x$, $y$, and $z$ axes of the detector's coordinate system.
        The true source location is shown as the red $\times$, and the GPSL-reconstructed position is shown as the black $+$.
        Voxels are $10$~cm cubes.
    }
    \label{fig:static_singles_cs137}
\end{figure}

To demonstrate static singles imaging, we placed point sources ${\sim}2$~m away from the imager at nominal angles of $0^\circ$ and $20^\circ$ from the imager's $-y$ axis, approximately level with the detector array center height, and collected data for up to ${\sim}10$~minutes.
We note that directly before this static measurement period, we moved the detector about the room for ${\sim}2$~minutes to perform a LiDAR scan, but we cut the trajectory and radiation data from this scan time out of the static singles analysis.
We then ran GPSL ($100$ iterations, no background) on the static portion of the dataset, reconstructing on all voxels ($10$~cm cubes) rather than only occupied voxels to more clearly show the utility of the detector response.
Results are given in Fig.~\ref{fig:static_singles_cs137} and show that GPSL reconstructs a cone of likelihoods or equivalently $z$-scores that is closely centered on the true source direction and that narrows in width as the measurement duration increases.
In the $20^\circ$ \cs case, a dwell time of even $5$~s is sufficient to determine the direction to the $175$~\uCi source to within $7^\circ$, despite such a large fraction of the reconstruction volume being included in the $2\sigma$ $z$-score contour.
With a dwell time of $180$~s, the $z$-score contours narrow considerably into a cone only a few degrees in opening angle, and the angular error reduces to $1.7^\circ$.
Although the most likely GPSL voxel does not provide an accurate source distance estimate as mentioned above, for completeness Fig.~\ref{fig:static_singles_cs137} also gives the GPSL-reconstructed activity at the true position, which is generally accurate to ${\sim}10\%$ in the \cs reconstructions shown.

\begin{figure}[!htbp]
    \centering
    \includegraphics[width=1.0\columnwidth]{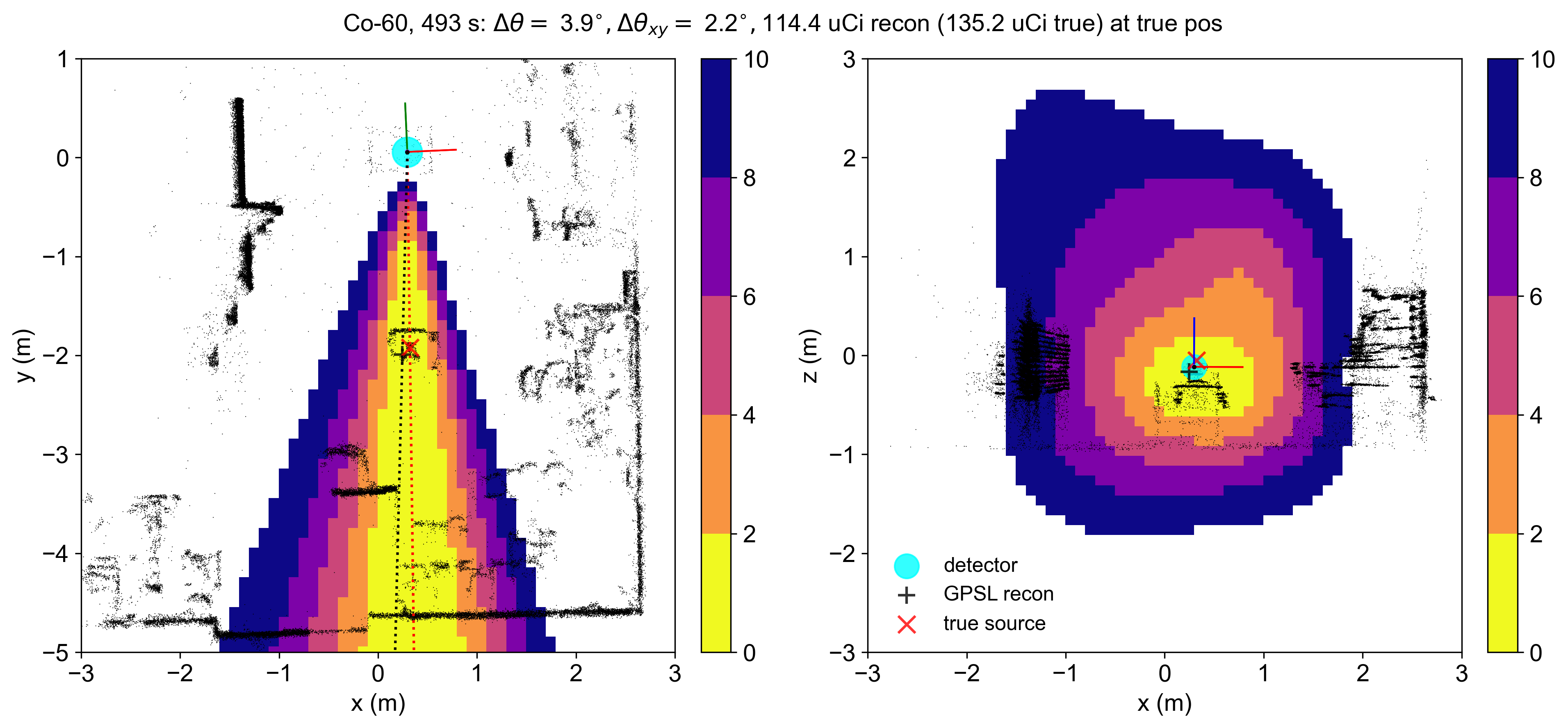}
    \includegraphics[width=1.0\columnwidth]{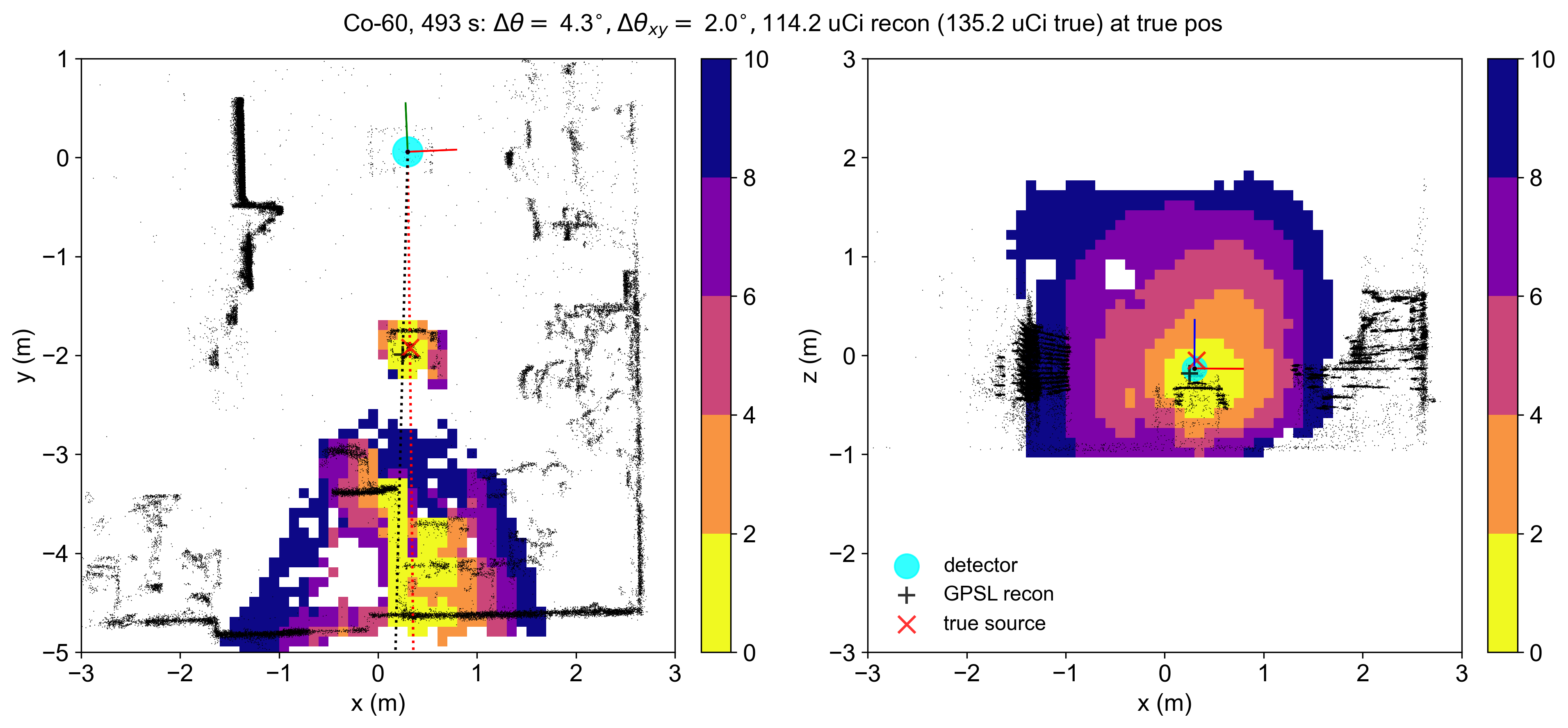}
    \caption{
        Top row: same as Fig.~\ref{fig:static_singles_cs137} but for a \co source in the nominal $0^\circ$ direction.
        Bottom row: same as top but with the LiDAR occupancy cut applied.
    }
    \label{fig:static_singles_co60}
\end{figure}

Similarly, Fig.~\ref{fig:static_singles_co60} shows an example \co ($1332$~keV) reconstruction with the source in the unrotated position for a longer dwell of $493$~s, in which GPSL determines the source angle to within $4^\circ$ of the true value.
In addition, while the static singles GPSL reconstruction may or may not on its own provide an accurate distance estimate, the LiDAR point cloud occupancy cut can of course break the $r^2$ degeneracy.
Applying the LiDAR cut to the \co data, the reconstructed source angle error remains ${\sim}4^\circ$ (largely due to voxel discretization), and the 3D position error is $22.5$~cm, i.e., only ${\sim}2$ voxel widths.
Similar position errors of ${\sim}2$ voxel widths in each direction ($37$~cm total) are obtained for the corresponding \cs reconstructions.

While angular reconstructions could reliably be obtained for \cs and \co, we note that the static singles results from \am at $59$~keV and of thermal neutrons from the moderated \cf source were less successful.
In particular, GPSL tended to reconstruct one or more cones of likelihood at angles far from the source direction.
We hypothesize that both these analyses were confounded by substantial non-point source emission behavior breaking the point source assumption of GPSL.
In the \cf case, there is likely ground scatter in the floor, while in the \am case, there is a substantial background (a combination of intrinsic and ambient) under the $59$~keV photopeak that cannot be separated out in the static singles analysis.
In the \am measurement approximately half of the \am ROI consisted of non-point-source background, whereas in the \cs and \co measurements, the photopeaks dominated the background.

As mentioned, GPSL in static singles mode did not (without the LiDAR occupancy cut) produce accurate distance and activity estimates, even at high statistics, despite this being theoretically possible with the omnidirectional multi-crystal response.
This is likely due to approximations in the detector response models, specifically the use of far-field (parallel beam) simulations to create responses independent of the source-to-detector distance.
Future work could develop multiple distance-dependent response functions and interpolate between them in order to more accurately capture near-field response effects.

Finally, we emphasize again that this static singles modality is enabled by the multi-crystal, active-masked design of the imager.
If we apply the isotropic or even monolithic response degradations of Section~\ref{sec:response_degradation} in this analysis, GPSL is unable break the $w / |\vec{r}|^2$ collinearity and the $|\vec{r}|^2$ degeneracy in Eq.~\ref{eq:lambda}, and is unable to provide a meaningful reconstruction.
The fact that an isotropic detector fails is intuitive, but the monolithic failure is perhaps more subtle---although a monolithic detector retains its angular anisotropy, it cannot determine where it was hit, and thus cannot correlate hits against its anisotropic efficiency to determine a source direction.

\subsection{Multiple point sources}\label{sec:apsl}

In this section, we demonstrate the imager's performance in reconstructing multiple point sources of the same nuclide within the same measured scene.
We placed three ${\sim}7$~\uCi \cs point sources in an indoor laboratory at least $3$~m apart from each other and performed a ${\sim}3$~minute free-moving survey of the room with the imager.

\begin{figure}[!htbp]
    \centering
    \includegraphics[width=0.8\columnwidth]{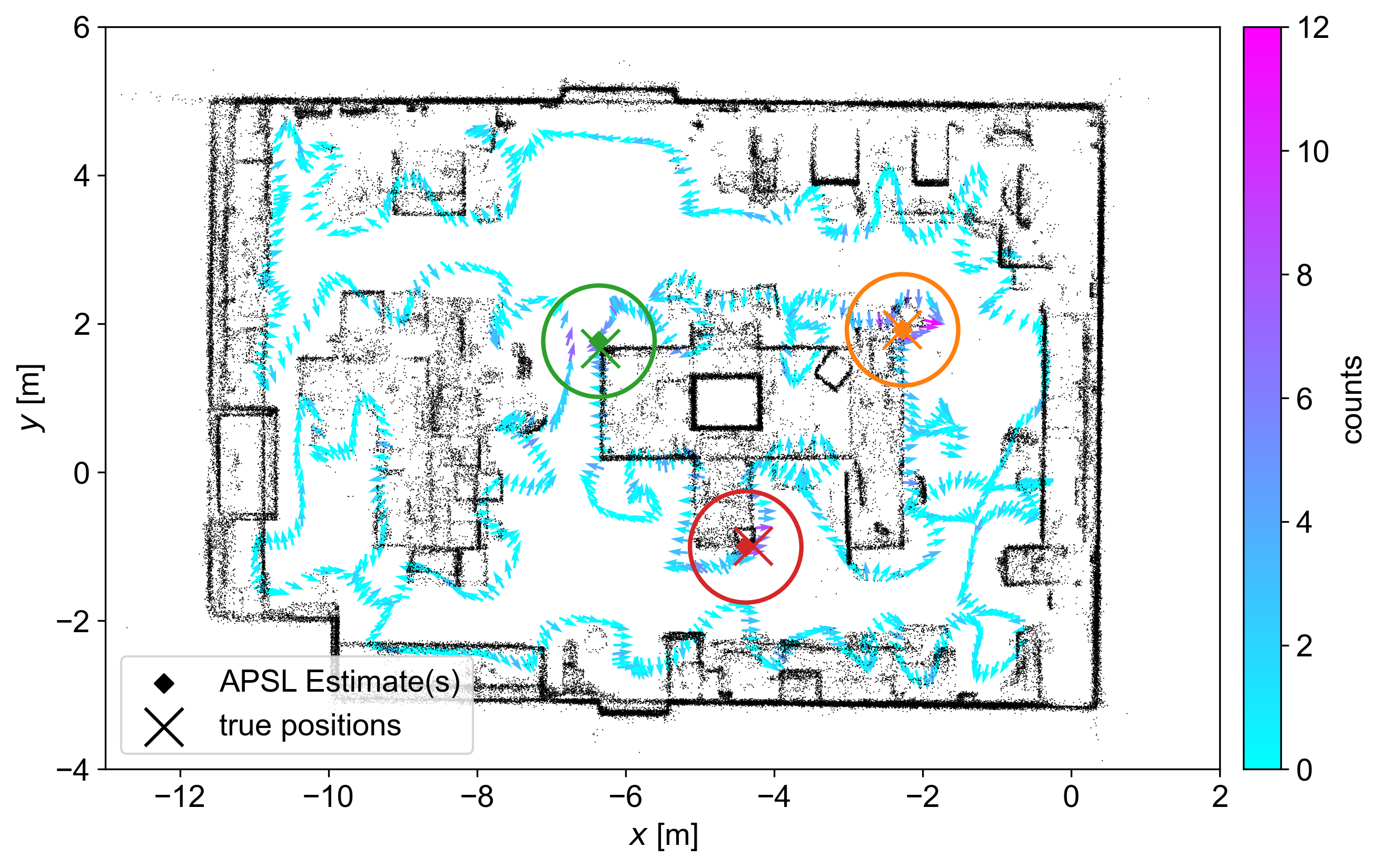}\\
    \includegraphics[width=0.8\columnwidth]{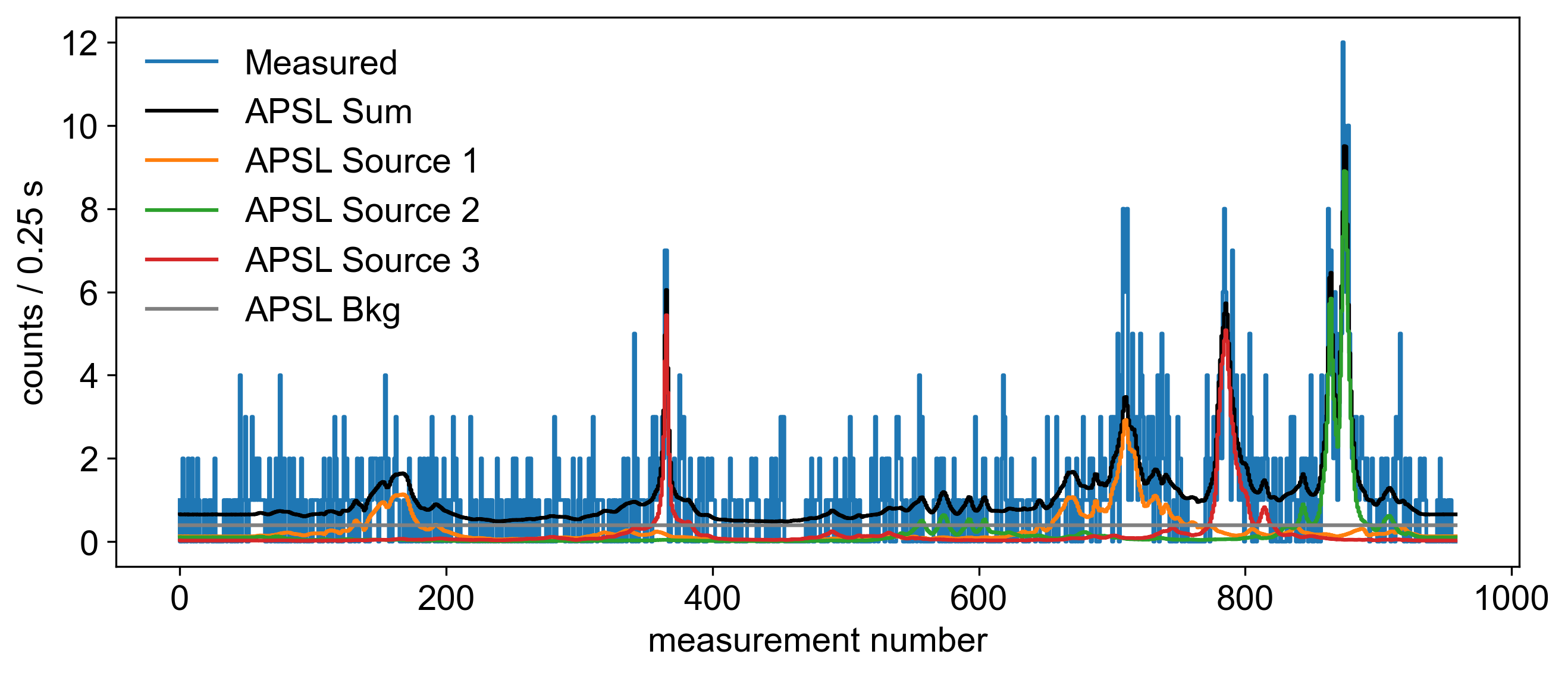}\\
    \includegraphics[width=0.8\columnwidth]{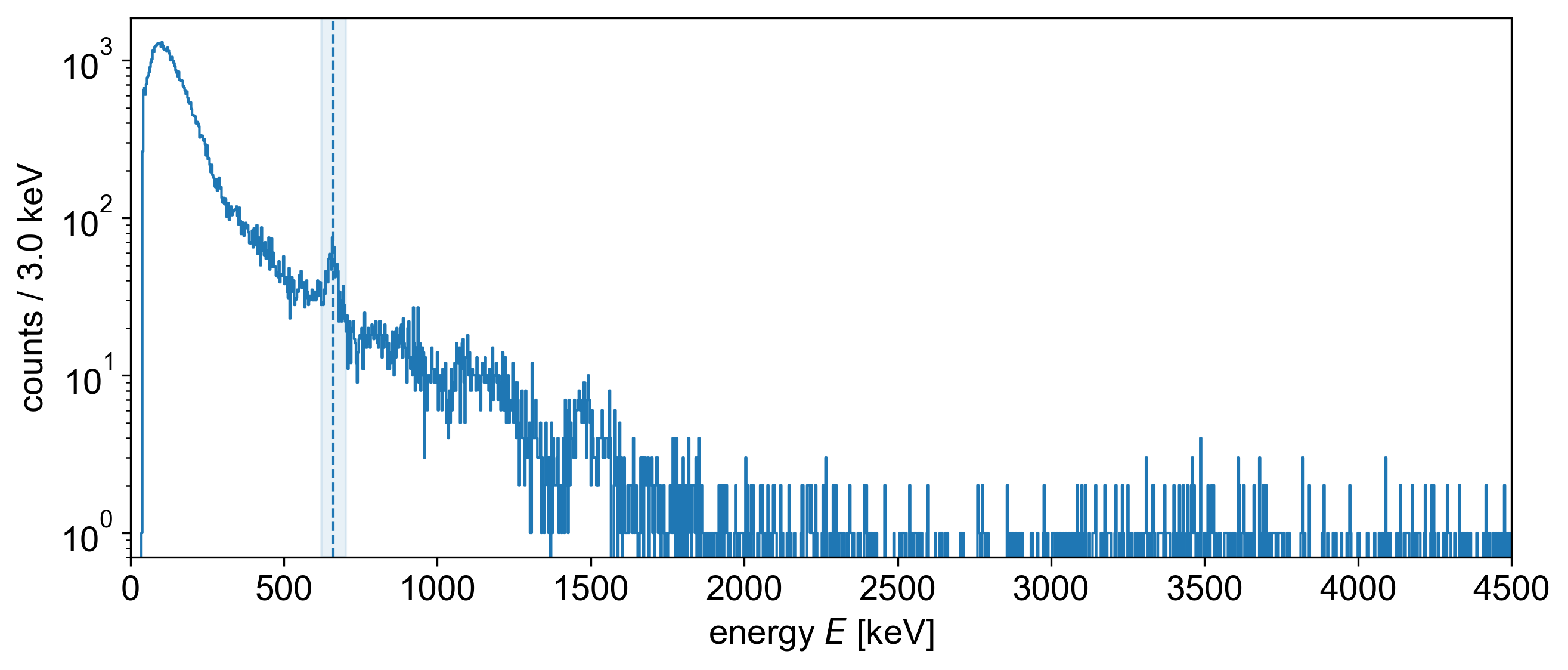}
    \caption{
        APSL reconstruction of three \cs point sources.
        Top: top-down view of the measurement and reconstruction.
        The detector trajectory is shown in the cyan-to-magenta arrows while the LiDAR point cloud (downsampled $200\times$, roof and floor removed) is shown in black dots.
        The reconstructed source positions are shown as orange, green, and red diamonds, while the ground truth positions are shown as the same color $\mathsf{X}$'s.
        The circles drawn around the reconstructed points are for visual clarity only and are not for instance indicative of any position uncertainty.
        Middle: ROI counts (summed over detectors) vs measurement number (i.e., time), showing the forward projection of each reconstructed source.
        Bottom: energy spectrum over the entire measurement, with the energy ROI of $661.7 \pm 40$~keV highlighted in blue.
    }
    \label{fig:apsl}
\end{figure}

Fig.~\ref{fig:apsl} shows a reconstruction of the measurement using Additive Point Source Localization (APSL)~\cite{hellfeld2019gamma, vavrek2020reconstructing}.
Even with the limited counts per measurement timestamp, the APSL reconstruction correctly identifies that three \cs sources are present and determines their locations with $xy$ distance errors $1$, $10$, and $10$~cm (and total distance errors of $8$, $19$, and $135$~cm) compared to the distances of closest approach to each true source location of $39$, $47$, and $47$~cm.
The largest total error ($135$~cm, red point in Fig.~\ref{fig:apsl}) arises from APSL placing the source location above instead of approximately level with it.
This degeneracy would likely be broken with better statistics (i.e., stronger sources or more time spent near the source) or if we used the multi-crystal response instead of summing over detectors (discussed further below).
Finally, the reconstructed activities are approximately correct but the two low-distance-error sources are biased low at $3.9$ and $4.5$~\uCi vs the expected $7.4$, $7.5$~\uCi, while the high-distance-error source is biased high at $10.5$~\uCi vs the expected $6.9$~\uCi.

We note that the LiDAR point cloud (black dots) provides context for the measured trajectory and the reconstructed source positions but does not directly inform the reconstruction---in APSL the reconstructed source positions are allowed to continuously vary in space and here are only constrained to a search region extending $2$~m beyond the trajectory bounds in each dimension.
The trajectory data was interpolated from $0.10$~s to $0.25$~s time bins to increase the counts per measurement timestep without overly sacrificing the spatial resolution of the trajectory.

In order to make this the APSL minimization computationally tractable, we had to sum over the $J=56$ active detectors (the ``monolithic'' degradation of Section~\ref{sec:response_degradation}).
As such, this demonstration relies primarily on proximity imaging, i.e., changes in the source-to-detector distance $|\vec{r}|$.
The monolithic reconstruction time was around $25$~s on a 2019 MacBook Pro with a $2.4$~GHz Intel Core i9 processor using 8 parallel processes for the APSL optimization, but the time required increases strongly with the number of detectors.
As shown in Ref.~\cite{vavrek2020reconstructing}, in these low count, monolithic detector scenarios, APSL may reconstruct the wrong number of sources and/or suffer from reduced accuracy.
However, in future analyses, the crystals could be summed into a small number of groups (perhaps by octant) rather than summed into a single monolith in order to retain some information on the gamma interaction location while keeping reconstruction times computationally tractable.

\subsection{Distributed sources}

\begin{figure}[!htbp]
    \centering
    \frame{\includegraphics[width=0.35\columnwidth]{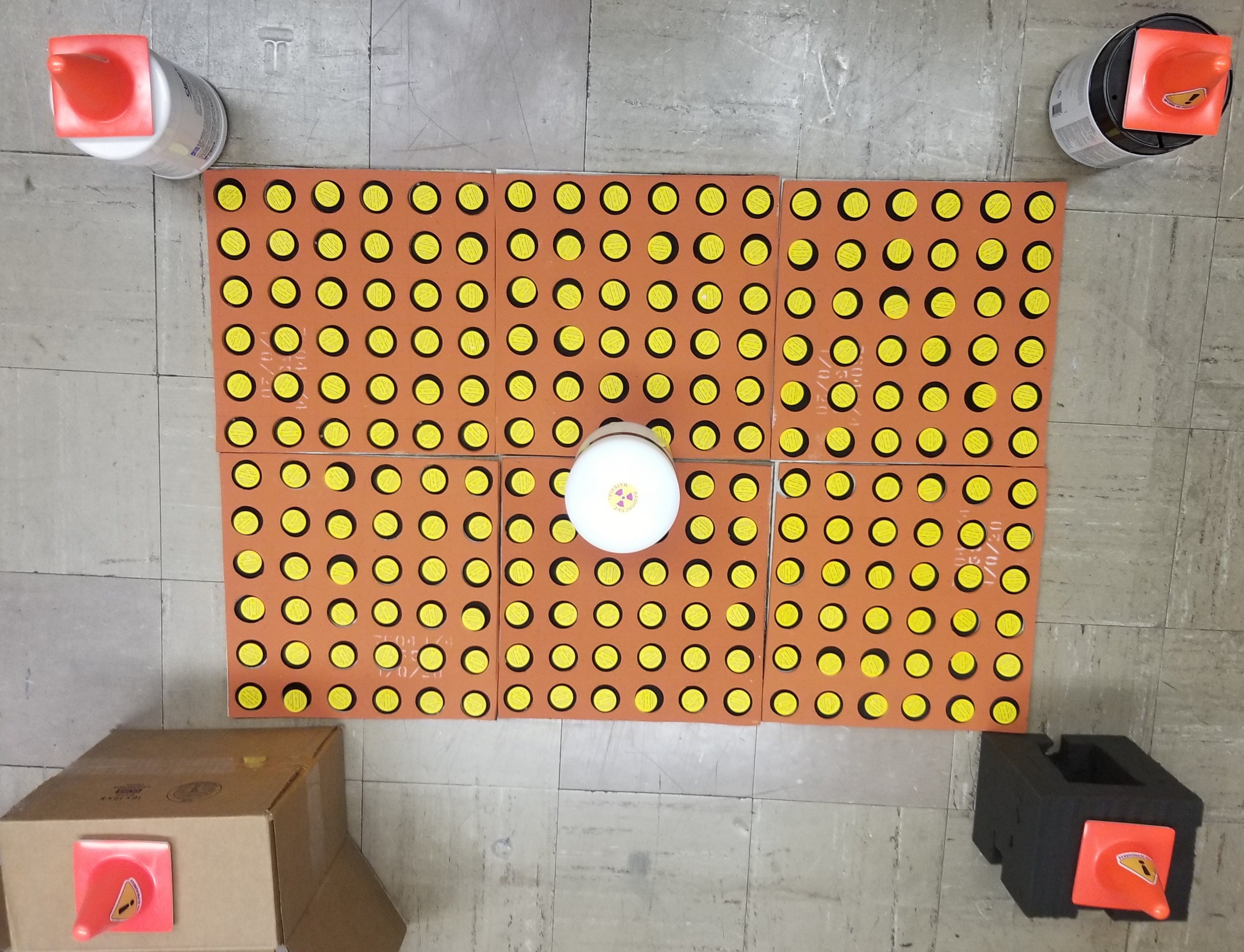}}
    \includegraphics[width=0.64\columnwidth]{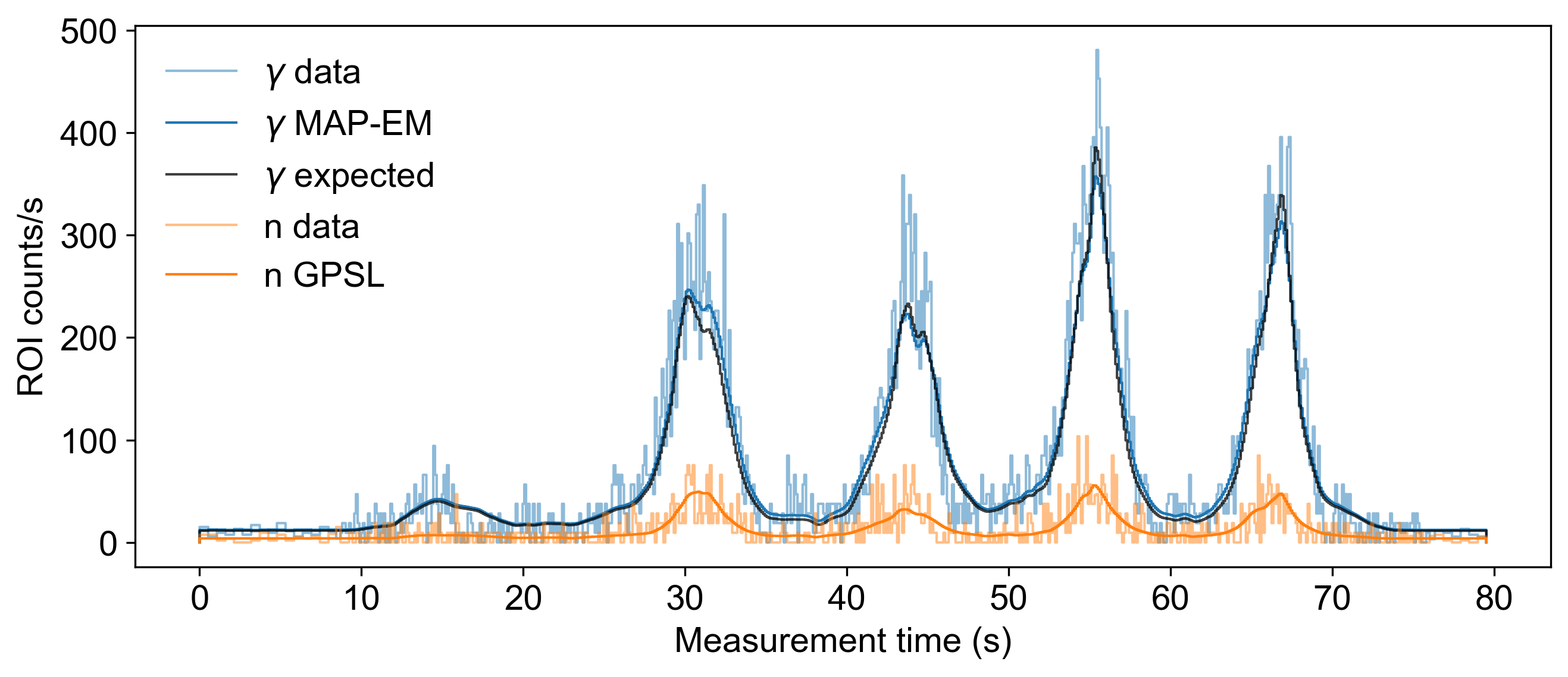}
    \includegraphics[width=1.00\columnwidth]{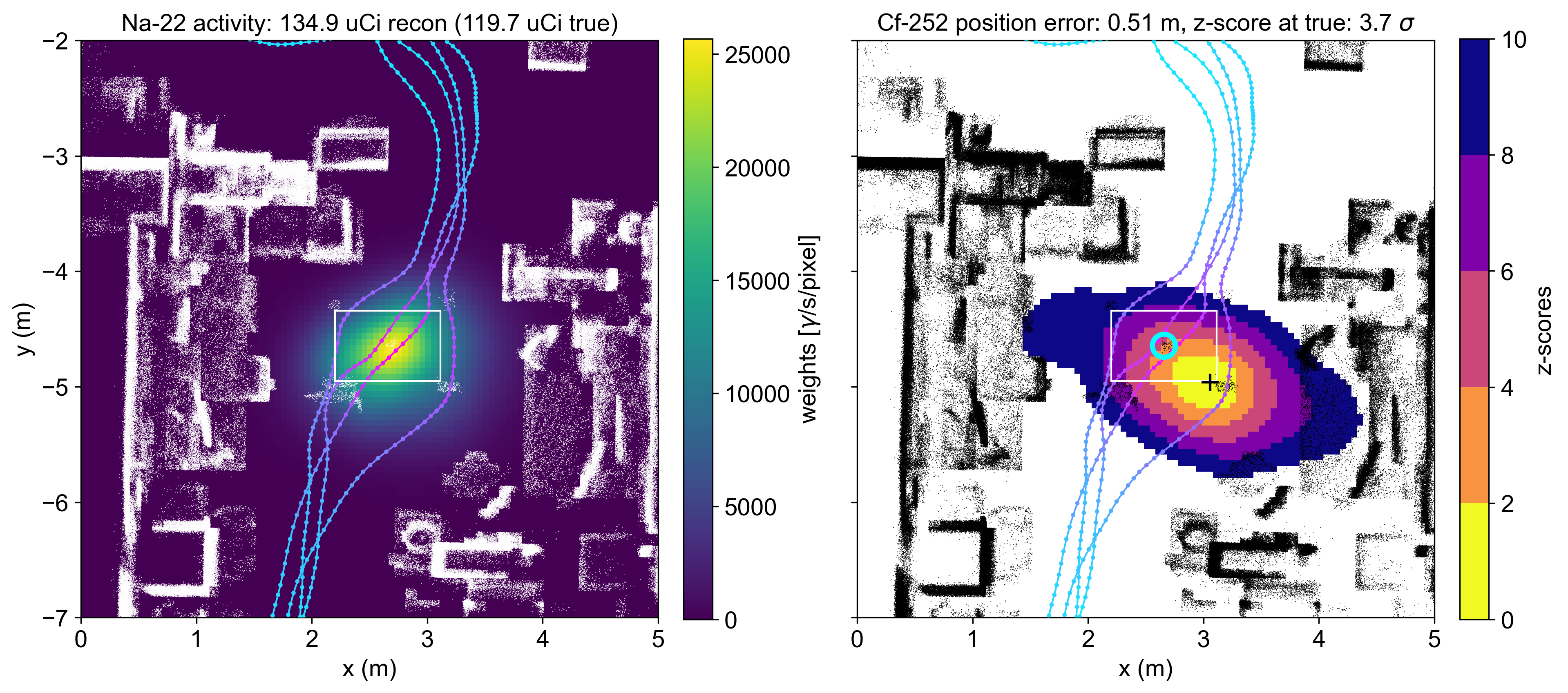}
    \caption{
        Simultaneous distributed gamma and point-like thermal neutron measurements and reconstructions.
        Top left: top-down photograph of the source, comprising a $2 \times 3$ arrangement of $6 \times 6$ panels of $0.55$~\uCi \na sources with a $42.6$~\uCi \cf source placed at the center.
        Top right: measured and reconstructed count rates in the gamma and neutron ROIs vs time.
        Bottom left: distributed gamma source reconstruction, showing MAP-EM reconstructed weights (photons/s/pixel) in the $511 \pm 40$~keV ROI.
        The white rectangle shows the approximate extent of the \na source panels.
        Bottom right: GPSL neutron source reconstruction.
        The blue circle identifies the \cf source in the LiDAR point cloud, while the black $+$ shows the most likely point source location determined by GPSL.
        Both reconstructions are constrained to the estimated ground plane rather than the entire LiDAR point cloud.
    }
    \label{fig:extended_sources}
\end{figure}

Finally, we demonstrate the imager's capability in distributed source reconstruction scenarios.
Because truly continuous distributed sources (e.g., in powdered form) can be difficult and hazardous to work with, we instead use a surrogate distributed source~\cite{vavrek2024surrogateI} consisting of six panels of $6 \times 6$ \na sealed point sources with a source grid pitch of $5.08$~cm.
Each source had a nominal activity of $1$~\uCi when manufactured in June~2020 and therefore had decayed for just under one $2.6$-year half-life to $0.55$~\uCi at the time of measurement in August~2022.
As an additional feature of the demonstration, we also placed a moderated $42.6$~\uCi \cf source in the center of the \na panels, which themselves were placed flat on the floor of an indoor laboratory environment---see Fig.~\ref{fig:extended_sources}.
Similar to the experiments of Ref.~\cite{hellfeld2021free}, this scenario could emulate a spill of (potentially mixed-nuclide) radioactive material that must be surveyed before cleanup and remediation.

An operator walked alongside and extended the imager over the combined \na + \cf source four times (twice in each direction), keeping the imager about $1.1$~m off the floor.
Care was taken to vary the $x$ coordinate of the trajectory over the source---and especially to pass over both the center and edges of the \na panels---in order to help break spatial degeneracies in the distributed source reconstruction.
Activities are fit to the floor of the 3D LiDAR model by defining the ground plane as a single layer of pixels at the height of the first percentile of all point cloud $z$ values.
We then perform two independent source reconstructions on this ground plane; for the \cf point source, we apply GPSL ($30$ iterations, $3400 \pm 500$~keV ROI) as above, and for the distributed \na we use MAP-EM ($50$ iterations, $511 \pm 40$~keV ROI) with an $L_{1/2}$ regularizer (strength $10^{-3}$).
Both reconstructions use a pixel size of $5$~cm.

The reconstruction results for the distributed \na and point \cf source are shown in Fig.~\ref{fig:extended_sources}.
The MAP-EM distributed source reconstruction of the \na gives a total reconstructed activity of $137.2$~\uCi compared to the true activity of $119.7$~\uCi, a discrepancy of $15\%$.
The reconstruction is well-centered within the true source extent (the white rectangle in Fig.~\ref{fig:extended_sources}), but approximately half of the activity resides outside this extent; this fraction will vary with the hyperparameters (number of iterations and $L_{1/2}$ strength).
The MAP-EM reconstructed photon count rates closely follow the expected count rates computed by forward-projecting the $216$ individual \na sources to the measured detector trajectory.
The GPSL reconstruction localizes the \cf source $57$~cm away from its true position, which is within the $3\,\sigma$ confidence level.
The reason for this bias is currently unknown, but may be due to un-modeled and potentially spatially-inhomogeneous neutron moderation in the room/subsurface, the human operator, and the foam block at the bottom right corner of the \na source nearest the GPSL-reconstructed \cf position.

It should be emphasized again that despite the apparent similarities of the MAP-EM and GPSL plots in Fig.~\ref{fig:extended_sources}, MAP-EM reconstructs distributed weights at a number of voxels, while GPSL assumes only one such pixel is ``active'' at a time and computes a likelihood and therefore $z$-score of that single-voxel model.
In addition, in a realistic radionuclide release scenario, the operator may not know (a) what nuclides are present and (b) whether to run distributed MAP-EM or point-like GPSL reconstructions.
To a large extent, (a) can be resolved (and potentially automated) by simply looking at the spectrum before deciding the peak ROI on which to reconstruct.
We note however that complex, multi-radionuclide spectra may present a challenge for the single-peak-ROI analysis presented here, and that full-spectral reconstructions are an area of ongoing research~\cite{lee2023simultaneous, xuGammarayEnergyimagingIntegrated2007, mehadjiExtensionListModeMLEM2018, chuAdvancedImagingAlgorithms2018}.
Issue (b) may be resolved by running both reconstruction methods and quantitatively comparing the goodness-of-fit of the two results, e.g., through the Akaike information criterion (AIC).

\section{Discussion}\label{sec:discussion}

We have demonstrated the multi-channel imager's utility in a variety of gamma and neutron mapping scenarios, ranging from free-moving handheld or vehicle-borne measurements designed to image primarily through $r^2$ modulation, to motion-constrained or static scenarios in which the imager's multi-crystal anisotropic omnidirectional response $\eta(\theta, \phi)$ plays a much larger role.
In general, the imager can localize and approximately quantify sources in measurement times on the order of one minute, with reconstructions requiring a similar or smaller amount of compute time.
As such, the imager is highly suitable for near-real-time operation in applications such as homeland security, contamination mapping, and nuclear decommissioning.

As shown in Tables~\ref{tab:gamma_validation} and \ref{tab:neutron_validation}, static dwell benchmarking studies of the absolute detector response show discrepancies of ${\sim}40\%$ for gammas and ${\sim}70\%$ for thermal neutrons.
Conversely, we note that some of the gamma activity predictions of Section~\ref{sec:response_degradation} are accurate to ${\lesssim}5\%$, and that some of the reconstructed gamma activities at the true positions in Section~\ref{sec:static_singles} are accurate to ${\lesssim}15\%$.
The reasons for the remaining discrepancies, especially the larger discrepancies in the dedicated benchmarking studies, are currently unknown, but here we discuss some possibilities.

The increase in discrepancy with photon energy in Table~\ref{tab:gamma_validation} is perhaps suggestive of a density error in CLLBC.
While we have used the manufacturer-specified $\rho = 4$~g/cm$^3$~\cite{rmd_cllbc_spec}, no more precision in the density value is available, nor is any information on density variability across crystals due to the crystal growth process.
By contrast, CLLBC crystals available from Berkeley Nucleonics Corporation have a quoted density of $\rho = 4.08$~g/cm$^3$~\cite{bnc_cllbc_spec}, and others have reported densities of $\rho = 4.13$~g/cm$^3$~\cite{guss2014scintillation}.
However, a $2$--$3\%$ density increase is incorrect in direction and insufficient in magnitude to fully account for these discrepancies.

Furthermore, the $8.25$~mm radial cut applied to model reduced light collection is a heuristic based on past efficiency characterizations of the previous-generation CLLBC-based imager demonstrated in Ref.~\cite{pavlovsky20193d}.
Such a radial cut, we note, is not present in the detailed CLLBC light collection simulations of Ref.~\cite{brown2023modelling}, but while the authors achieve good agreement in spectral shapes, they do not report absolute efficiency comparisons.
Without an accurate idea of how this reduced light collection volume should be modelled, if at all, improved absolute efficiency validation will be challenging.
Destructive tests and collimated efficiency scans (perhaps even at the intra-crystal level) could be valuable in this regard, but very labor-intensive.

There remain several research directions that could be explored in future work.
First, all analyses in this work have used singles-mode data, but the quasi-random pattern of detector modules was also designed to enable high-performance Compton imaging (as shown for CZT in Ref.~\cite{hellfeld2021free}).
It would be interesting for example to compare system performance under Compton imaging performance (with lower statistics but better angular resolution) against the static singles imaging results of Section~\ref{sec:static_singles} (with better statistics but worse angular resolution).
Second, given that CLLBC is sensitive to both thermal and fast neutrons, it could be valuable to develop fast neutron imaging and a rough neutron spectral unfolding method in order to enable neutron dose estimation.
Third, we have recently developed a framework for analyzing and optimizing the resolution vs efficiency tradeoff as variable-performance segments of highly-segmented detectors are accumulated~\cite{aversano2023data}.
Given the $50$--$60$ active detector segments in the present imager, it would be interesting to apply this framework to optimizing the imager's spectral performance for various measurement tasks.
Finally, all the measurements in this work were controlled by a human operator, but it would be valuable to develop autonomous path-planning algorithms that can best leverage the system's imaging capability.

\section{Conclusion}
We have demonstrated a new CLLBC-based gamma- and neutron-sensitive imaging detector capable of both free-moving and static measurements.
The imager shows good quantitative performance in localizing and determining the strength of both gamma and neutron sources, and should be useful in a wide variety of applications including homeland security, contamination mapping, and nuclear decommissioning.

\section*{Author contributions}
Conceptualization, RP, DH, JWC; Methodology, JRV, RP, VN, BJQ, JWC; Software, JRV, RP, DH, THYJ; Validation, JRV, JWC; Formal Analysis, JRV; Investigation, JRV, RP, BJQ, JWC; Resources, VN, JWC; Data Curation, JRV, JWC; Writing – Original Draft Preparation, JRV, JWC; Writing – Review \& Editing, JRV, VN, DH, THYJ, BJQ, JWC; Visualization, JRV, JWC; Supervision, JWC; Project Administration, JWC; Funding Acquisition, JWC.



\section*{Funding}
This material is based upon work supported by the Defense Threat Reduction Agency under IAA numbers 10027-28022, 10027-23334, and 13081-36242.
This support does not constitute an express or implied endorsement on the part of the United States Government.
Distribution A: approved for public release, distribution is unlimited.

This document was prepared as an account of work sponsored by the United States Government. While this document is believed to contain correct information, neither the United States Government nor any agency thereof, nor the Regents of the University of California, nor any of their employees, makes any warranty, express or implied, or assumes any legal responsibility for the accuracy, completeness, or usefulness of any information, apparatus, product, or process disclosed, or represents that its use would not infringe privately owned rights. Reference herein to any specific commercial product, process, or service by its trade name, trademark, manufacturer, or otherwise, does not necessarily constitute or imply its endorsement, recommendation, or favoring by the United States Government or any agency thereof, or the Regents of the University of California. The views and opinions of authors expressed herein do not necessarily state or reflect those of the United States Government or any agency thereof or the Regents of the University of California.

This manuscript has been authored by an author at Lawrence Berkeley National Laboratory under Contract No.~DE-AC02-05CH11231 with the U.S.~Department of Energy. The U.S.~Government retains, and the publisher, by accepting the article for publication, acknowledges, that the U.S.~Government retains a non-exclusive, paid-up, irrevocable, world-wide license to publish or reproduce the published form of this manuscript, or allow others to do so, for U.S.~Government purposes.

This research used the Lawrencium computational cluster resource provided by the IT Division at the Lawrence Berkeley National Laboratory (Supported by the Director, Office of Science, Office of Basic Energy Sciences, of the U.S. Department of Energy under Contract No. DE-AC02-05CH11231).

\section*{Acknowledgments}
The authors thank Micah Folsom and Bryan van der Ende for useful discussions on Geant4, Weronika Wolszczak for useful discussions on scintillator materials, Marco Salathe for assistance with data curation and tuning SLAM parameters, and Andy Haefner of Gamma Reality Inc.\ for driving the Gator vehicle in the measurements of Fig.~\ref{fig:gator_drive-bys}.

\section*{Conflicts of interest}
The authors declare no conflict of interest.
The sponsors had no role in the design, execution, interpretation, or writing of the study, or in the decision to publish the results.

\bibliographystyle{IEEEtran}
\bibliography{biblio}

\end{document}

%% file: gamma_validation.tex
\begin{table}[!h]
\centering
\caption{Laboratory gamma validation results}
\label{tab:gamma_validation}
\begin{tabular}{llrrr}
\toprule
source & direction & $\eta_\text{meas}$ [cm$^2$] & $\eta_\text{sim}$ [cm$^2$] & $\eta_\text{sim}/\eta_\text{meas}$ \\
\midrule
Cs-137 & left & $2.58 \pm 0.04$ & $3.66$ & $1.42 \pm 0.02$ \\
Cs-137 & back & $2.58 \pm 0.04$ & $3.63$ & $1.41 \pm 0.02$ \\
Cs-137 & front & $1.39 \pm 0.03$ & $2.72$ & $1.96 \pm 0.04$ \\
Cs-137 & right & $2.49 \pm 0.04$ & $3.66$ & $1.47 \pm 0.02$ \\
Cs-137 & top & $2.39 \pm 0.04$ & $3.46$ & $1.44 \pm 0.02$ \\
Co-60 & left & $0.80 \pm 0.02$ & $1.22$ & $1.53 \pm 0.04$ \\
Am-241 & left & $18.77 \pm 0.86$ & $23.27$ & $1.24 \pm 0.06$ \\
\bottomrule
\end{tabular}
\end{table}

%% file: neutron_validation.tex
\begin{table}[!h]
\centering
\caption{Outdoor thermal neutron validation results}
\label{tab:neutron_validation}
\begin{tabular}{lrrr}
\toprule
direction & sim rate & meas rate &       sim/meas \\
\midrule
front     &  $(5.77 \pm 0.18) \times 10^{-7}$ &  $(3.31 \pm 0.09) \times 10^{-7}$ &  $1.75 \pm 0.07$ \\
back      &  $(6.01 \pm 0.21) \times 10^{-7}$ &  $(3.51 \pm 0.11) \times 10^{-7}$ &  $1.71 \pm 0.08$ \\
left      &  $(6.30 \pm 0.24) \times 10^{-7}$ &  $(3.78 \pm 0.11) \times 10^{-7}$ &  $1.67 \pm 0.08$ \\
right     &  $(6.32 \pm 0.23) \times 10^{-7}$ &  $(3.63 \pm 0.11) \times 10^{-7}$ &  $1.74 \pm 0.08$ \\
bottom    &  $(1.68 \pm 0.07) \times 10^{-6}$ &  $(9.96 \pm 0.14) \times 10^{-7}$ &  $1.68 \pm 0.07$ \\
\bottomrule
\end{tabular}
\end{table}

%% file: main.bbl
\begin{thebibliography}{10}
\providecommand{\url}[1]{#1}
\csname url@samestyle\endcsname
\providecommand{\newblock}{\relax}
\providecommand{\bibinfo}[2]{#2}
\providecommand{\BIBentrySTDinterwordspacing}{\spaceskip=0pt\relax}
\providecommand{\BIBentryALTinterwordstretchfactor}{4}
\providecommand{\BIBentryALTinterwordspacing}{\spaceskip=\fontdimen2\font plus
\BIBentryALTinterwordstretchfactor\fontdimen3\font minus
  \fontdimen4\font\relax}
\providecommand{\BIBforeignlanguage}[2]{{%
\expandafter\ifx\csname l@#1\endcsname\relax
\typeout{** WARNING: IEEEtran.bst: No hyphenation pattern has been}%
\typeout{** loaded for the language `#1'. Using the pattern for}%
\typeout{** the default language instead.}%
\else
\language=\csname l@#1\endcsname
\fi
#2}}
\providecommand{\BIBdecl}{\relax}
\BIBdecl

\bibitem{vetter2019advances}
K.~Vetter, R.~Barnowski, J.~W. Cates, A.~Haefner, T.~H. Joshi, R.~Pavlovsky,
  and B.~J. Quiter, ``Advances in nuclear radiation sensing: Enabling {3-D}
  gamma-ray vision,'' \emph{Sensors}, vol.~19, no.~11, p. 2541, 2019.

\bibitem{haefner2017handheld}
A.~Haefner, R.~Barnowski, P.~Luke, M.~Amman, and K.~Vetter, ``Handheld
  real-time volumetric {3-D} gamma-ray imaging,'' \emph{Nuclear Instruments and
  Methods in Physics Research Section A: Accelerators, Spectrometers, Detectors
  and Associated Equipment}, vol. 857, pp. 42--49, 2017.

\bibitem{hellfeld2021free}
D.~Hellfeld, M.~S. Bandstra, J.~R. Vavrek, D.~L. Gunter, J.~C. Curtis,
  M.~Salathe, R.~Pavlovsky, V.~Negut, P.~J. Barton, J.~W. Cates \emph{et~al.},
  ``Free-moving quantitative gamma-ray imaging,'' \emph{Scientific reports},
  vol.~11, no.~1, pp. 1--14, 2021.

\bibitem{vavrek2024surrogateIII}
J.~R. Vavrek, J.~Lee, M.~Salathe, M.~S. Bandstra, D.~Hellfeld, B.~J. Quiter,
  and T.~H. Joshi, ``Surrogate distributed radiological sources {III}:
  quantitative distributed source reconstructions,'' \emph{arXiv preprint
  arXiv:2412.02926}, 2024.

\bibitem{vavrek2023free}
J.~R. Vavrek, B.~J. Quiter, D.~Hellfeld, J.~Szornel, R.~Pavlovsky,
  H.~Gonzalez-Raymat, H.~Wainwright, and C.~Eddy-Dilek, ``Free-moving mapping
  of radioactive contaminants at the {Savannah River Site F-Area} wetlands,''
  in \emph{Waste Management Symposium Conference Proceedings}, 2023.

\bibitem{bandstra2016radmap}
M.~S. Bandstra, T.~J. Aucott, E.~Brubaker, D.~H. Chivers, R.~J. Cooper, J.~C.
  Curtis, J.~R. Davis, T.~H. Joshi, J.~Kua, R.~Meyer \emph{et~al.}, ``{RadMAP}:
  The radiological multi-sensor analysis platform,'' \emph{Nuclear Instruments
  and Methods in Physics Research Section A: Accelerators, Spectrometers,
  Detectors and Associated Equipment}, vol. 840, pp. 59--68, 2016.

\bibitem{salathe2021determining}
M.~Salathe, B.~Quiter, M.~Bandstra, J.~Curtis, R.~Meyer, and C.~Chow,
  ``Determining urban material activities with a vehicle-based multi-sensor
  system,'' \emph{Physical Review Research}, vol.~3, no.~2, p. 023070, 2021.

\bibitem{joshi2017measurement}
T.~H. Joshi, B.~J. Quiter, J.~S. Maltz, M.~S. Bandstra, A.~Haefner,
  N.~Eikmeier, E.~Wagner, T.~Luke, R.~Malchow, and K.~McCall, ``Measurement of
  the energy-dependent angular response of the {ARES} detector system and
  application to aerial imaging,'' \emph{IEEE Transactions on Nuclear Science},
  vol.~64, no.~7, pp. 1754--1760, 2017.

\bibitem{bandstra2022mapping}
M.~S. Bandstra, D.~Hellfeld, J.~Lee, B.~J. Quiter, M.~Salathe, J.~R. Vavrek,
  and T.~H. Joshi, ``Mapping the minimum detectable activities of gamma-ray
  sources in a {3-D} scene,'' \emph{IEEE Transactions on Nuclear Science},
  2022.

\bibitem{bandstra2021improved}
M.~S. Bandstra, D.~Hellfeld, J.~R. Vavrek, B.~J. Quiter, K.~Meehan, P.~J.
  Barton, J.~W. Cates, A.~Moran, V.~Negut, R.~Pavlovsky \emph{et~al.},
  ``Improved gamma-ray point source quantification in three dimensions by
  modeling attenuation in the scene,'' \emph{IEEE Transactions on Nuclear
  Science}, vol.~68, no.~11, pp. 2637--2646, 2021.

\bibitem{okabe2024tetris}
R.~Okabe, S.~Xue, J.~R. Vavrek, J.~Yu, R.~Pavlovsky, V.~Negut, B.~J. Quiter,
  J.~W. Cates, T.~Liu, B.~Forget \emph{et~al.}, ``Tetris-inspired detector with
  neural network for radiation mapping,'' \emph{Nature Communications},
  vol.~15, no.~1, p. 3061, 2024.

\bibitem{boardman2020single}
D.~Boardman, A.~Sarbutt, A.~Flynn, and M.~Guenette, ``Single pixel compressive
  gamma-ray imaging with randomly encoded masks,'' \emph{Journal of
  Instrumentation}, vol.~15, no.~04, p. P04014, 2020.

\bibitem{ghelman2024wide}
M.~Ghelman, N.~Kopeika, S.~Rotman, N.~B. David, E.~Vax, and A.~Osovizky, ``Wide
  energetic response of $4\pi$ directional gamma detector based on combination
  of {Compton} scattering and photoelectric effect,'' \emph{IEEE Transactions
  on Nuclear Science}, 2024.

\bibitem{kemp2023real}
S.~Kemp, S.~Kumar, C.~Bakker, and J.~Rogers, ``Real-time radiological source
  term estimation for multiple sources in cluttered environments,'' \emph{IEEE
  Transactions on Nuclear Science}, 2023.

\bibitem{kitayama2022feasibility}
Y.~Kitayama, M.~Nogami, and K.~Hitomi, ``Feasibility study on a gamma-ray
  imaging using three-dimensional shadows of gamma rays,'' IEEE NSS-MIC, 2022.

\bibitem{hu2023wide}
Y.~Hu, Z.~Lyu, P.~Fan, T.~Xu, S.~Wang, Y.~Liu, and T.~Ma, ``A wide energy range
  and 4$\pi$-view gamma camera with interspaced position-sensitive scintillator
  array and embedded heavy metal bars,'' \emph{Sensors}, vol.~23, no.~2, p.
  953, 2023.

\bibitem{rossi2023gamma}
F.~Rossi, L.~Cosentino, F.~Longhitano, S.~Minutoli, P.~Musico, M.~Osipenko,
  G.~E. Poma, M.~Ripani, and P.~Finocchiaro, ``The gamma and neutron sensor
  system for rapid dose rate mapping in the {CLEANDEM} project,''
  \emph{Sensors}, vol.~23, no.~9, p. 4210, 2023.

\bibitem{steinberger2020imaging}
W.~M. Steinberger, M.~L. Ruch, N.~Giha, A.~D. Fulvio, P.~Marleau, S.~D. Clarke,
  and S.~A. Pozzi, ``Imaging special nuclear material using a handheld dual
  particle imager,'' \emph{Scientific reports}, vol.~10, no.~1, p. 1855, 2020.

\bibitem{lopez2022neutron}
R.~Lopez, W.~Steinberger, N.~Giha, P.~Marleau, S.~Clarke, and S.~Pozzi,
  ``Neutron and gamma imaging using an organic glass scintillator handheld dual
  particle imager,'' \emph{Nuclear Instruments and Methods in Physics Research
  Section A: Accelerators, Spectrometers, Detectors and Associated Equipment},
  vol. 1042, p. 167407, 2022.

\bibitem{sinclair2014silicon}
L.~Sinclair, P.~Saull, D.~Hanna, H.~Seywerd, A.~MacLeod, and P.~Boyle,
  ``Silicon photomultiplier-based compton telescope for safety and security
  {(SCoTSS)},'' \emph{IEEE Transactions on Nuclear Science}, vol.~61, no.~5,
  pp. 2745--2752, 2014.

\bibitem{sinclair2020end}
L.~E. Sinclair, A.~McCann, P.~R. Saull, R.~L. Mantifel, C.~V. Ouellet, P.-L.
  Drouin, A.~M. Macleod, B.~Le~Gros, I.~Summerell, J.~H. Hovgaard
  \emph{et~al.}, ``End-user experience with the {SCoTSS Compton} imager and
  directional survey spectrometer,'' \emph{Nuclear Instruments and Methods in
  Physics Research Section A: Accelerators, Spectrometers, Detectors and
  Associated Equipment}, vol. 954, p. 161683, 2020.

\bibitem{murtha2021tomographic}
N.~Murtha, L.~Sinclair, P.~Saull, A.~McCann, and A.~MacLeod, ``Tomographic
  reconstruction of a spatially-extended source from the perimeter of a
  restricted-access zone using a {SCoTSS} compton gamma imager,'' \emph{Journal
  of Environmental Radioactivity}, vol. 240, p. 106758, 2021.

\bibitem{gri_lamp}
``{GRI-LAMP}: Real-time {3D} radiation mapping,'' retrieved Oct.~22, 2023 from
  \url{https://www.gammareality.com/lamp}.

\bibitem{h3d_products}
``{H3D} products,'' retrieved Oct.~22, 2023 from
  \url{https://h3dgamma.com/ProductTiles.php}.

\bibitem{Pavlovsky2018}
R.~Pavlovsky \emph{et~al.}, ``{3-D} radiation mapping in real-time with the
  {Localization and Mapping Platform (LAMP)} from unmanned aerial systems and
  man-portable configurations,'' \emph{arXiv:1901.05038}, 2018.

\bibitem{Durrant-Whyte2006_1}
H.~Durrant-Whyte and T.~Bailey, ``{Simultaneous Localization and Mapping: Part
  I},'' \emph{\href{https://ieeexplore.ieee.org/document/1638022}{IEEE Robot.
  Autom. Mag.}}, vol.~13, no.~2, 2006.

\bibitem{Durrant-Whyte2006_2}
------, ``{Simultaneous Localization and Mapping: Part II},''
  \emph{\href{https://ieeexplore.ieee.org/document/1678144}{IEEE Robot. Autom.
  Mag.}}, vol.~13, no.~3, 2006.

\bibitem{Hess2016}
W.~Hess, D.~Kohler, H.~Rapp, and D.~Andor, ``Real-time loop closure in {2D
  LIDAR SLAM},'' \emph{\href{https://ieeexplore.ieee.org/document/7487258}{in
  Proc. IEEE International Conference on Robotics and Automation}}, pp.
  1271--1278, 2016.

\bibitem{shepp1982maximum}
L.~A. Shepp and Y.~Vardi, ``Maximum likelihood reconstruction for emission
  tomography,'' \emph{{IEEE Trans. on Medical Imaging}}, vol.~1, no.~2, pp.
  113--122, 1982.

\bibitem{hellfeld2019gamma}
D.~Hellfeld, T.~H. Joshi, M.~S. Bandstra, R.~J. Cooper, B.~J. Quiter, and
  K.~Vetter, ``Gamma-ray point-source localization and sparse image
  reconstruction using {Poisson} likelihood,'' \emph{IEEE Transactions on
  Nuclear Science}, vol.~66, no.~9, pp. 2088--2099, 2019.

\bibitem{vavrek2020reconstructing}
J.~R. Vavrek, D.~Hellfeld, M.~S. Bandstra, V.~Negut, K.~Meehan, W.~J.
  Vanderlip, J.~W. Cates, R.~Pavlovsky, B.~J. Quiter, R.~J. Cooper
  \emph{et~al.}, ``Reconstructing the position and intensity of multiple
  gamma-ray point sources with a sparse parametric algorithm,'' \emph{IEEE
  Transactions on Nuclear Science}, vol.~67, no.~11, pp. 2421--2430, 2020.

\bibitem{agostinelli2003geant4}
S.~Agostinelli, J.~Allison, K.~Amako, J.~Apostolakis, H.~Araujo, P.~Arce,
  M.~Asai, D.~Axen, S.~Banerjee, G.~Barrand \emph{et~al.}, ``Geant4—a
  simulation toolkit,'' \emph{Nuclear Instruments and Methods in Physics
  Research Section A: Accelerators, Spectrometers, Detectors and Associated
  Equipment}, vol. 506, no.~3, pp. 250--303, 2003.

\bibitem{allison2006geant4}
J.~Allison, K.~Amako, J.~Apostolakis, H.~Araujo, P.~A. Dubois, M.~Asai,
  G.~Barrand, R.~Capra, S.~Chauvie, R.~Chytracek \emph{et~al.}, ``Geant4
  developments and applications,'' \emph{IEEE Transactions on Nuclear Science},
  vol.~53, no.~1, pp. 270--278, 2006.

\bibitem{allison2016recent}
J.~Allison, K.~Amako, J.~Apostolakis, P.~Arce, M.~Asai, T.~Aso, E.~Bagli,
  A.~Bagulya, S.~Banerjee, G.~Barrand \emph{et~al.}, ``{Recent developments in
  Geant4},'' \emph{Nuclear Instruments and Methods in Physics Research Section
  A: Accelerators, Spectrometers, Detectors and Associated Equipment}, vol.
  835, pp. 186--225, 2016.

\bibitem{rmd_cllbc_spec}
\BIBentryALTinterwordspacing
``{CLLBC} gamma-neutron scintillator properties,'' accessed October 24, 2022.
  [Online]. Available:
  \url{https://www.dynasil.com//assets/CLLBC-Gamma-Neutron-Scintillator-Properties.pdf}
\BIBentrySTDinterwordspacing

\bibitem{guss2014scintillation}
P.~P. Guss, T.~G. Stampahar, S.~Mukhopadhyay, A.~Barzilov, and A.~Guckes,
  ``Scintillation properties of a {Cs2LiLa(Br6)90\%(Cl6)10\%:Ce3+ (CLLBC)}
  crystal,'' in \emph{Radiation Detectors: Systems and Applications XV}, vol.
  9215.\hskip 1em plus 0.5em minus 0.4em\relax International Society for Optics
  and Photonics, 2014, p. 921505.

\bibitem{knoll2010radiation}
G.~F. Knoll, \emph{Radiation detection and measurement}.\hskip 1em plus 0.5em
  minus 0.4em\relax John Wiley \& Sons, 2010.

\bibitem{pavlovsky2019miniprism}
R.~Pavlovsky \emph{et~al.}, ``{MiniPRISM: 3D} realtime gamma-ray mapping from
  small unmanned aerial systems and handheld scenarios,'' IEEE NSS-MIC
  Conference Record, 2019.

\bibitem{farzanehpoor2021feasibility}
A.~Farzanehpoor~Alwars and F.~Rahmani, ``A feasibility study of gamma ray
  source finder development for multiple sources scenario based on a {Monte
  Carlo} simulation,'' \emph{Scientific Reports}, vol.~11, no.~1, p. 6121,
  2021.

\bibitem{pavlovsky20193d}
R.~Pavlovsky, J.~W. Cates, W.~J. Vanderlip, T.~H.~Y. Joshi, A.~Haefner,
  E.~Suzuki, R.~Barnowski, V.~Negut, A.~Moran, K.~Vetter, and B.~J. Quiter,
  ``{3D} gamma-ray and neutron mapping in real-time with the {Localization and
  Mapping Platform} from unmanned aerial systems and man-portable
  configurations,'' \emph{arXiv:1908.06114}, 2019.

\bibitem{lingenfelter2009sparsity}
D.~J. Lingenfelter, J.~A. Fessler, and Z.~He, ``Sparsity regularization for
  image reconstruction with {Poisson} data,'' in \emph{Computational Imaging
  VII}, vol. 7246.\hskip 1em plus 0.5em minus 0.4em\relax SPIE, 2009, pp.
  96--105.

\bibitem{xu2010lhalf}
Z.~Xu, H.~Zhang, Y.~Wang, X.~Chang, and Y.~Liang, ``{$L_{1/2}$
  regularization},'' \emph{Science China Information Sciences}, vol.~53, no.~6,
  pp. 1159--1169, 2010.

\bibitem{greiff2021gamma}
M.~Greiff, E.~Rofors, A.~Robertsson, R.~Johansson, and R.~Tyllstr{\"o}m,
  ``Gamma-ray imaging with spatially continuous intensity statistics,'' in
  \emph{2021 IEEE/RSJ International Conference on Intelligent Robots and
  Systems (IROS)}.\hskip 1em plus 0.5em minus 0.4em\relax IEEE, 2021, pp.
  5257--5262.

\bibitem{bissantz2008statistical}
N.~Bissantz, B.~A. Mair, and A.~Munk, ``A statistical stopping rule for {MLEM}
  reconstructions in {PET},'' in \emph{2008 IEEE Nuclear Science Symposium
  Conference Record}.\hskip 1em plus 0.5em minus 0.4em\relax IEEE, 2008, pp.
  4198--4200.

\bibitem{montgomery2020novel}
L.~Montgomery, A.~Landry, G.~Al~Makdessi, F.~Mathew, and J.~Kildea, ``A novel
  {MLEM} stopping criterion for unfolding neutron fluence spectra in radiation
  therapy,'' \emph{Nuclear Instruments and Methods in Physics Research Section
  A: Accelerators, Spectrometers, Detectors and Associated Equipment}, vol.
  957, p. 163400, 2020.

\bibitem{ben2013heuristic}
F.~Ben~Bouall{\`e}gue, J.-F. Crouzet, and D.~Mariano-Goulart, ``A heuristic
  statistical stopping rule for iterative reconstruction in emission
  tomography,'' \emph{Annals of nuclear medicine}, vol.~27, pp. 84--95, 2013.

\bibitem{joshi2020multi}
T.~Joshi, B.~Quiter, J.~Curtis, M.~Bandstra, R.~Cooper, D.~Hellfeld,
  M.~Salathe, A.~Moran, and J.~Vavrek, ``Multi-modal free-moving data fusion
  ({MFDF}) v1.0,'' Lawrence Berkeley National Laboratory (LBNL), Berkeley, CA
  (United States), Tech. Rep., 2020.

\bibitem{joshi2021radkit}
T.~Joshi, B.~Quiter, J.~Curtis, M.~Folsom, M.~Bandstra, N.~Abgrall, R.~Cooper,
  D.~Hellfeld, M.~Salathe, A.~Moran \emph{et~al.}, ``radkit v1.2,'' Lawrence
  Berkeley National Laboratory (LBNL), Berkeley, CA (United States), Tech.
  Rep., 2021.

\bibitem{hellfeld2021radkit}
D.~Hellfeld, J.~C. Curtis, M.~Salathe, M.~S. Bandstra, M.~Folsom, J.~R. Vavrek,
  A.~Moran, B.~J. Quiter, R.~J. Cooper, and T.~H.~Y. Joshi, ``Radkit: Python
  packages for radiological and contextual data processing and fusion,'' IEEE
  NSS-MIC, 2021.

\bibitem{gorski2005healpix}
K.~M. G{\'o}rski, E.~Hivon, A.~J. Banday, B.~D. Wandelt, F.~K. Hansen,
  M.~Reinecke, and M.~Bartelmann, ``{HEALPix: A framework for high-resolution
  discretization and fast analysis of data distributed on the sphere},''
  \emph{The Astrophysical Journal}, vol. 622, no.~2, p. 759, 2005.

\bibitem{Zonca2019}
\BIBentryALTinterwordspacing
A.~Zonca, L.~Singer, D.~Lenz, M.~Reinecke, C.~Rosset, E.~Hivon, and K.~Gorski,
  ``healpy: equal area pixelization and spherical harmonics transforms for data
  on the sphere in {Python},'' \emph{Journal of Open Source Software}, vol.~4,
  no.~35, p. 1298, Mar. 2019. [Online]. Available:
  \url{https://doi.org/10.21105/joss.01298}
\BIBentrySTDinterwordspacing

\bibitem{chytracek2006geometry}
R.~Chytracek, J.~McCormick, W.~Pokorski, and G.~Santin, ``Geometry description
  markup language for physics simulation and analysis applications,''
  \emph{IEEE Transactions on Nuclear Science}, vol.~53, no.~5, pp. 2892--2896,
  2006.

\bibitem{geant4_application_book}
``{GEANT4} book for application developers, 10.7,'' 2020, retrieved Sept.~23,
  2021, from
  \url{https://geant4-userdoc.web.cern.ch/UsersGuides/ForApplicationDeveloper/fo/BookForApplicationDevelopers.pdf}.

\bibitem{shin2014geant4}
J.~W. Shin, S.-W. Hong, S.-I. Bak, D.~Y. Kim, and C.~Y. Kim, ``{GEANT4} and
  {PHITS} simulations of the shielding of neutrons from the {252 Cf} source,''
  \emph{Journal of the Korean Physical Society}, vol.~65, pp. 591--598, 2014.

\bibitem{vavrek2024surrogateII}
J.~R. Vavrek, C.~C. Hines, M.~S. Bandstra, D.~Hellfeld, M.~A. Heine, Z.~M.
  Heiden, N.~R. Mann, B.~J. Quiter, and T.~H. Joshi, ``Surrogate distributed
  radiological sources {II}: aerial measurement campaign,'' \emph{IEEE
  Transactions on Nuclear Science}, 2024.

\bibitem{bukartas2022accuracy}
A.~Bukartas, J.~Wallin, R.~Finck, and C.~R{\"a}{\"a}f, ``Accuracy of a
  {Bayesian} technique to estimate position and activity of orphan gamma-ray
  sources by mobile gamma spectrometry: Influence of imprecisions in
  positioning systems and computational approximations,'' \emph{{PLoS ONE}},
  vol.~17, no.~6, p. e0268556, 2022.

\bibitem{zhao2022two}
Q.~Zhao, Z.~Wang, L.~Li, X.~Li, C.~Zhao, M.~Zhao, F.~Li, M.~Cheng, B.~Zhu,
  R.~Zhou \emph{et~al.}, ``A two-dimensional array detector for determining the
  direction to gamma-ray source,'' \emph{Nuclear Instruments and Methods in
  Physics Research Section A: Accelerators, Spectrometers, Detectors and
  Associated Equipment}, vol. 1040, p. 166985, 2022.

\bibitem{vavrek2024surrogateI}
J.~R. Vavrek, M.~S. Bandstra, D.~Hellfeld, B.~J. Quiter, and T.~H. Joshi,
  ``Surrogate distributed radiological sources {I}: point-source array design
  methods,'' \emph{IEEE Transactions on Nuclear Science}, 2024.

\bibitem{lee2023simultaneous}
J.~Lee, M.~Bandstra, B.~Quiter, D.~Gunter, and K.~Vetter, ``Simultaneous
  radiological spectral decomposition and source mapping with a single
  detector,'' in \emph{2023 {IEEE} Nuclear Science Symposium and Medical
  Imaging Conference Record ({NSS}/{MIC})}, 2023.

\bibitem{xuGammarayEnergyimagingIntegrated2007}
\BIBentryALTinterwordspacing
D.~Xu and Z.~He, ``\BIBforeignlanguage{en}{Gamma-ray energy-imaging integrated
  spectral deconvolution},'' \emph{\BIBforeignlanguage{en}{Nuclear Instruments
  and Methods in Physics Research Section A: Accelerators, Spectrometers,
  Detectors and Associated Equipment}}, vol. 574, no.~1, pp. 98--109, Apr 2007.
  [Online]. Available:
  \url{https://www.sciencedirect.com/science/article/pii/S0168900207002549}
\BIBentrySTDinterwordspacing

\bibitem{mehadjiExtensionListModeMLEM2018}
B.~Mehadji, M.~Dupont, Y.~Boursier, and C.~Morel, ``Extension of the list-mode
  {MLEM} algorithm for poly-energetic imaging with a {Compton} camera,'' in
  \emph{2018 {IEEE} Nuclear Science Symposium and Medical Imaging Conference
  Proceedings ({NSS}/{MIC})}, Nov 2018, pp. 1--8.

\bibitem{chuAdvancedImagingAlgorithms2018}
\BIBentryALTinterwordspacing
J.~Chu, ``Advanced imaging algorithms with position-sensitive gamma-ray
  detectors,'' Thesis, University of Michigan, 2018. [Online]. Available:
  \url{http://deepblue.lib.umich.edu/handle/2027.42/146036}
\BIBentrySTDinterwordspacing

\bibitem{bnc_cllbc_spec}
\BIBentryALTinterwordspacing
``High resolution dual mode {CLLBC} scintillators,'' accessed July 16, 2021.
  [Online]. Available:
  \url{https://www.berkeleynucleonics.com/sites/default/files/products/datasheets/cllbc-ds-11-27-18\_0.pdf}
\BIBentrySTDinterwordspacing

\bibitem{brown2023modelling}
J.~Brown, L.~Chartier, D.~Boardman, J.~Barnes, and A.~Flynn, ``Modelling the
  response of {CLLBC (Ce) and TLYC (Ce) SiPM}-based radiation detectors in
  mixed radiation fields with {Geant4},'' \emph{arXiv preprint
  arXiv:2303.09709}, 2023.

\bibitem{aversano2023data}
G.~Aversano, H.~Parrilla, and J.~R. Vavrek, ``Data-driven event selection in
  pixelated cadmium zinc telluride {(CZT)} detectors for improved gamma-ray
  spectrometry,'' in \emph{Proceedings of the {INMM}}, 2023.

\end{thebibliography}
